\pdfoutput=1
\documentclass[a4paper,fleqn,usenatbib]{mnras}
\usepackage{amsmath}
\usepackage{tablefootnote}
\usepackage{natbib}
\usepackage{bm}
\usepackage{graphicx}
\usepackage[T1]{fontenc}
\usepackage{lmodern}
\usepackage{amssymb}
\usepackage{lipsum}
\usepackage{flushend}
\makeatletter

\patchcmd{\NAT@citex}
  {\@citea\NAT@hyper@{%
     \NAT@nmfmt{\NAT@nm}%
     \hyper@natlinkbreak{\NAT@aysep\NAT@spacechar}{\@citeb\@extra@b@citeb}%
     \NAT@date}}
  {\@citea\NAT@nmfmt{\NAT@nm}%
   \NAT@aysep\NAT@spacechar\NAT@hyper@{\NAT@date}}{}{}

\patchcmd{\NAT@citex}
  {\@citea\NAT@hyper@{%
     \NAT@nmfmt{\NAT@nm}%
     \hyper@natlinkbreak{\NAT@spacechar\NAT@@open\if*#1*\else#1\NAT@spacechar\fi}%
       {\@citeb\@extra@b@citeb}%
     \NAT@date}}
  {\@citea\NAT@nmfmt{\NAT@nm}%
   \NAT@spacechar\NAT@@open\if*#1*\else#1\NAT@spacechar\fi\NAT@hyper@{\NAT@date}}
  {}{}

\makeatother

\defcitealias{lazarian2000velocity}{LP00}
\defcitealias{lazarian2004velocity}{LP04}
\defcitealias{lazarian2006studying}{LP06}
\defcitealias{lazarian2008studying}{LP08}
\defcitealias{lazarian2012statistical}{LP12}
\defcitealias{0004-637X-818-2-178}{LP16}
\defcitealias{kandel2016extending}{KLP}
\defcitealias{goldreich1995toward}{GS95}
\defcitealias{esquivel2011velocity}{EL11}
\defcitealias{burkhart2014measuring}{BX14}
\defcitealias{zhuravleva2012constraints}{ZX12}
\defcitealias{dickman1985largescale}{DK85}
\defcitealias{lazarian2003statistics}{LE03}
\title[Study of velocity centroids]{Study of velocity centroids based on the theory of fluctuations in position-position-velocity space}

\author[Kandel, Lazarian \& Pogosyan]{
D. Kandel,$^{1,3}$\thanks{E-mail: dkandel@ualberta.ca}
A. Lazarian$^{2}$
and D. Pogosyan$^{1,3}$
\\
$^{1}$Physics Department, University of Alberta, Edmonton, T6G 2E1, Canada\\
$^{2}$Department of Astronomy, University of Wisconsin, 475 North Charter Street, Madison, WI 53706, USA\\
$^{3}$CNRS and UPMC, UMR 7095, Institut d'Astrophysique de Paris, F-75014, Paris, France\\
}

\date{\today}

\pubyear{2016}

\begin{document}
\label{firstpage}
\pagerange{\pageref{firstpage}--\pageref{lastpage}}
\maketitle

\begin{abstract}
We study possibility of obtaining velocity spectra by studying turbulence in an optically thick medium using velocity centroids (VCs). We find that the regime of universal, i.e. independent of underlying turbulence statistics, fluctuations discovered originally within the velocity channel analysis (VCA) carries over to the statistics of VCs. In other words, for large absorptions the VC lose their ability to reflect the spectra of turbulence. Combining our present study with the earlier studies of centroids in Esquivel \& Lazarian, we conclude that centroids are applicable for studies subsonic/transsonic turbulence for the range of scales that is limited by the  absorption effects. We also consider VCs based on absorption lines and define the range of their applicability. We address the problem of analytical description of spectra and anisotropies of fluctuations that are available through studies using VC. We obtain spectra and anisotropy of VC fluctuations arising from Alfv\'en, slow and fast modes that constitute the compressible MHD cascade to address the issue of anisotropy of VC statistics, and show how the VC anisotropy can be used to find the media magnetization as well as to identify and separate contributions from Alfv\'en, slow and fast modes. Our study demonstrates that VCs are complementary to the tools provided by the VCA. In order to study turbulent volume for which the resolution of single dish telescopes is insufficient, we demonstrate how the studies of anisotropy can be performed using interferometers. 
\end{abstract}

\section{Introduction}
The interstellar medium (ISM) is magnetized and turbulent.  Observations of non-thermal Doppler broadening of spectral lines, fluctuations of density and synchrotron emission (see reviews by \citealt{cho2003mhd}; \citealt{elmegreen2004interstellar}, \citealt{mac2004control}; \citealt{ballesteros2007molecular}; \citealt{2007ARA&A..45..565M}; \citealt{2009SSRv..143..357L}) are some   cases that suggest the ubiquity of MHD turbulence in the ISM. Moreover, the estimated Reynolds number in the ISM is of the order $10^8$, which corresponds to turbulent motions (\citealt{1949ApJ...110..329C}). Thus, MHD turbulence is of key importance for star formation (\citealt{federrath2012star}; \citealt{federrath2013origin};  \citealt{salim2015universal}), propagation and acceleration of cosmic rays and other fundamental astrophysical processes (see \citealt{brandenburg2013astrophysical} and references therein). 

Understanding the interstellar turbulence requires one to successfully study the statistics of underlying turbulent field, in particular to obtain the velocity and density spectrum of the turbulent field. One of the frequently used techniques to obtain velocity spectra using spectral lines is  velocity centroids (VC), which are first moments of spectral line (see \citealt{munch1958space}; \citealt{kleiner1985large}; \citealt{o1987evidence}; \citealt{miesch1999velocity}). \citet{esquivel2005velocity} studied the extent up to which VCs correctly reflect the turbulence velocity spectrum.  Numerical studies (see \citealt{esquivel2005velocity}; \citealt{esquivel2007statistics}) have later shown that the spectrum can be obtained adequately correctly for subsonic turbulence, but is significantly distorted for supersonic turbulence. The former property of VC was found to be complementary to the properties of the two more recent techniques based on the analytical description of turbulence, namely, Velocity Channel Analysis (VCA, \citealt{lazarian2000velocity, lazarian2004velocity}, hereafter \citetalias{lazarian2000velocity} and \citetalias{lazarian2004velocity}, respectively)  and Velocity Coordinate Spectrum (VCS, \citealt{lazarian2006studying, lazarian2008studying}, hereafter \citetalias{lazarian2006studying} and \citetalias{lazarian2008studying}, respectively) . The findings of those papers suggest that studies of subsonic turbulence are possible only for heavier species moving with the flow, e.g. heavy ions, atoms and molecules moving together with atomic or molecular hydrogen. The analytical description that was at the core of the VCA and VCS techniques made them advantageous compared to VCs, which were studied only numerically.

The VCA technique has been successfully tested and elaborated in a number of subsequent papers (\citealt{lazarian2001emissivity}; \citealt{chepurnov2009turbulence}; \citealt{burkhart2013turbulence}) and  was successfully applied to a number of observations (see an incomplete list in \citealt{2009SSRv..143..357L}).  In terms of the spectra study, the VCA technique suggests a way of disentangling velocity and density contributions to the channel maps through varying the thickness of the corresponding maps. This technique has been successfully applied to HI and CO data in e.g. \cite{padoan2009power}, \cite{chepurnov2010extending} and \cite{chepurnov2015turbulence} to obtain velocity spectra. 

Velocity and density spectra do not provide a complete description of an underlying turbulent field, particularly in a magnetized plasma. MHD turbulence is known to be anisotropic  with magnetic field defining the direction of anisotropy (\citealt{montgomery1981anisotropic};  \citealt{shebalin1983anisotropy}; \citealt{higdon1984density}). In this regard, apart from spectra, the direction of magnetic field and magnetisation can also be studied with the PPV data. The first demonstration of this possibility of obtaining magnetic field direction was given in \citet{lazarian2002seeing}, which was followed by the subsequent empirical studies in this direction (e.g. \citealt{esquivel2011velocity}; \citealt{burkhart2014measuring}). The research based on analytical studies was carried out not with spectroscopic, but synchrotron data.  \citet[hereafter \citetalias{lazarian2012statistical} and \citetalias{0004-637X-818-2-178}, respectively]{lazarian2012statistical, 0004-637X-818-2-178} carried out the analytical description of synchrotron intensity and polarization fluctuation for the model of turbulence containing Alfv\'en, fast and slow mode cascades (see \citealt{brandenburg2013astrophysical} and references therein). The analytical description of anisotropies in \citetalias{lazarian2012statistical} and \citetalias{0004-637X-818-2-178} allowed to relate the analytical predictions of magnetization of the media and the observed anisotropy of the synchrotron polarization and intensity fluctuations. The approach in these works was used in \citet[hereafter \citetalias{kandel2016extending}]{kandel2016extending} to provide the analytical description of the anisotropies in PPV space, extending the VCA in the case of anisotropic turbulence. The study in \citetalias{kandel2016extending} provides the foundation for our present study of anisotropy using VC. 

In this paper, we aim to present a comprehensive study of velocity centroids by extending centroids to study turbulence in a self-absorbing medium, as well as with absorption lines. We also present a formulation of centroids anisotropy in order to study anisotropy of an underlying MHD turbulence. The structure of our paper is as follows: in Sec. \ref{sec:centroid} we review important expressions of centroids through position-position-velocity (PPV) space formalism, which we extend in Sec. \ref{othick} to describe the centroid statistics in the presence of self-absorption. In Sec. \ref{abline}, we describe the usefulness and limitations of centroids in the case of absorption lines. In Sec. \ref{sec:genanisotropy}, we develop a general formalism to study turbulence anisotropies through centroids, and we apply this formalism to specific MHD modes in Sec. \ref{sec:mhdmodes}. In Sec. \ref{sec:interfero}, we show how interferometric data can be used to study turbulence using centroids technique. The major findings of the paper are summarised in Sec. \ref{sec:majorfinding}, and comparisons between VCA and centroids is made in Sec. \ref{vcacentroid}. The practical issues of turbulence studies using centroids is presented in Sec. \ref{finiteres}. In Sec. \ref{discuss} we discuss about the foundation of our technique, the assumptions we used and the range of applicability of centroids; and relate our study with other existing  technique. Finally, we summarise our important findings in Sec. \ref{summ}.

\section{Velocity centroids}\label{sec:1}
The in situ point-wise measurements in XYZ space are not available with spectroscopic measurements. Therefore,  measurements of the intensity of emissions are defined in PPV space or XYV volume, where the turbulence information along $z$-axis is subject to a non-linear transformation due to the mapping to the line-of-sight (LOS) velocity axis. Doppler shifts are affected only by the LOS (which we assume to be aligned with $z$-axis) component of turbulent velocities, which to simplify our notations we denote as $v$ (instead of $v_z$).

The theory of intensity fluctuations in PPV space was pioneered in \citetalias{lazarian2000velocity} and was later extended for special cases in \citetalias{lazarian2004velocity}, \citetalias{lazarian2006studying}, \citetalias{lazarian2008studying} as well
as in the anisotropy studies in \citetalias{kandel2016extending}. In this paper, we work in PPV space to study how the VCs reflect velocity spectra as well as  anisotropic nature of the velocity and density statistics in magnetized turbulence. In this section, we begin to develop framework to study centroids in the presence of self-absorption. For that we first derive the centroid correlation as well as structure function through PPV space. The description of centroids in PPV space turns out to be valuable to understand the effects of self-absorption on centroids. 

\begin{table*}
\centering
\caption{Different types of centroids}
\begin{tabular}{l c c}  
\hline  
\emph{Type of centroids} & \emph{Definition} & \emph{Structure function}\\
\hline
Normalised centroid & $C_N(\bm{X})=\frac{\int \mathop{\mathrm{d}v_1}v_1I_{v1}(\bm{X})}{\int \mathop{\mathrm{d}v_1}I_{v1}(\bm{X})}$ & Not used in this paper\\
Unnormalised centroid & $C(\bm{X})=\int \mathop{\mathrm{d}v_1}v_1I_{v1}(\bm{X})$ & Eq. \eqref{correlation}\\
Modified centroids & only structure function is defined  & Eq. \eqref{modifiedcorrelation}\\
Restricted Centroids$^{\dagger}$ & Eq. \eqref{sharpdef} & Eq. \eqref{evenodd}\\
\hline
\label{tab:centroidstype}
\end{tabular}
\begin{tabular}{l}
$^{\dagger}$ Restricted centroids are used for saturated absorption lines.
\end{tabular}
\end{table*}

Before starting formal derivations, we assume turbulence to be homogeneous and isotropic in Sec. \ref{sec:1} and Sec. \ref{abline}, while the assumption of isotropy will be relaxed in Sec. \ref{sec:genanisotropy} and Sec. \ref{sec:mhdmodes} to study anisotropies due to magnetisation of a turbulent media. Furthermore, we assume that the statistical descriptors of velocity and density fluctuations obey power-law behaviour on scale, i.e $\propto r^{-p}$.  Fluctuations whose correlations are dominated by large scales are called {\it steep}, while those whose correlations are dominated by small scales are called {\it shallow}. Formally, $p<0$ characterises steep statistics, and for such statistics it is usually convenient to use structure function. Shallow statistics is characterised by $p>0$ and it is convenient to use correlation function to describe such statistics. Our derivation is based for steep velocity spectra, for which we assume line of sight projected structure function to be of the form $D_z(r)\sim r^{-\nu}$. For shallow density spectra we use $\xi_\rho(r)\sim \langle\rho\rangle^2+\langle\delta\rho^2\rangle(r/r_c)^{-\nu_\rho}$, while for steep spectra we use $\xi_\rho(r)\sim \langle\rho\rangle^2+\langle\delta\rho^2\rangle-\langle\delta\rho^2\rangle(r/r_c)^{-\nu_\rho}$, where $r_c$ is the correlation length of the density field. In the case of shallow spectra, the dynamical range exists at $r>r_c$, for steep spectra, the dynamical range exists at $r<r_c$.

The centroid in PPV space is the moment of intensity defined as (see \citealt{miesch1994statistical})
\begin{equation}\label{eq7}
C_N(\bm{X})=\frac{\int \mathop{\mathrm{d}v_1}v_1I_{v1}(\bm{X})}{\int \mathop{\mathrm{d}v_1}I_{v1}(\bm{X})}~,
\end{equation}
where $I_{v1}$ is the spectral intensity and $v_1$ is the line of sight velocity. By solving  one dimensional radiative transfer equation in the case of self-absorbing emission in spectral lines, one can obtain the spectral intensity as (see \citetalias{lazarian2004velocity})
\begin{equation}\label{eq8}
I_v(\bm{X})=\frac{\epsilon}{\alpha}\left[1-\mathrm{e}^{-\alpha\rho_s(\bm{X},v)}\right]~,
\end{equation}
where $\epsilon$ is the emissivity coefficient, $\alpha$ is the self-absorption coefficient, $v$ is the LOS velocity, $\bm{X}$ is the sky projected two dimensional position vector, $\rho_s$ is the PPV space density given by
\begin{equation}\label{eq:rhos}
\rho_s(\bm{X},v)=\int\mathop{\mathrm{d}z}\rho(\bm{x})\Phi(v-u(\bm{x}))~,
\end{equation}
with $\bm{x}$ is the three-dimensional position vector of a turbulent point, $\Phi(v-u(\bm{x}))$ being the Maxwell's distribution of the thermal component of LOS velocity, and $u(\bm{x})$ being the LOS turbulent velocity. 

Due to presence of denominator term in the definition of centroids in Eq. \eqref{eq7}, $C_M(\bm{X})$ is non-linear function of $I$, which  complicates relation between statistics of centroids and intensity. To remedy this difficulty, LE03 introduced `unnormalised' velocity centroids (UVC) defined as
\begin{equation}\label{eq:unnormalised}
C(\bm{X})=\int \mathop{\mathrm{d}v_1}v_1I_{v1}(\bm{X})~.
\end{equation}
A summary of different types of centroids is presented in Table \ref{tab:centroidstype}. In all the subsequent sections, we carry out analysis with unnormalised centroids.
\begin{table*}
\centering
\caption{List of notations used in this paper.}
\begin{tabular}{l c c}  
\hline  
\emph{Parameter} & \emph{Meaning} & \emph{First appearance}\\
\hline
$\bm{x}$ & 3-D position vector & Eq. \eqref{eq:rhos}\\
$\bm{X}$ & 2-D position vector & Eq. \eqref{eq7}\\
$\bm{r}$ & 3-D separation $\bm{x}_2-\bm{x}_1$ &  Eq. \eqref{eq:firstcorr} \\ 
$\bm{R}$ & 2-D separation $\bm{X}_2-\bm{X}_1$ &  Eq. \eqref{eq:firstcorr} \\ 
$\rho_s(\bm{X},v)$ & Density of emitters in the PPV space & Eq. \eqref{eq8} \\
$\Phi(v)$ & Maxwell's distribution function & Eq. \eqref{eq:rhos}\\
$\beta_{\text{T}}$ & Thermal broadening & Eq. \eqref{centroidcorrm}\\
$\xi_\rho(\bm{x})$ & Density correlation function & Eq. \eqref{eq:firstcorr}\\
$D_z(\bm{r})$ & LOS projected velocity structure function & Eq. \eqref{centroidcorrm}\\
$d_\rho(\bm{r})$ & Density structure function & Eq. \eqref{correlation}\\
$S$ & Size of a turbulent cloud & Eq. \eqref{eq:firstcorr}\\
$v_{\text{ab}}$ & Velocity cut-off introduced by self-absorption & Eq. \eqref{window}\\
$\tau(\bm{X},v)$ & Optical depth & Eq. \eqref{eq:opticaldepth}\\
$\Delta$ & Characteristic width of the window in the absorption line study & Eq. \eqref{sharpdef}\\
$\hat{\lambda}$ & Direction of the mean magnetic field &  Eq. \eqref{fouriercorr}\\ 
$a_{\bm{k}}$ & Random amplitude of a mode & Eq. \eqref{fouriercorr}\\
$\hat{\xi}_{\bm{k}}$ & Direction of allowed displacement in a plasma & Eq. \eqref{fouriercorr}\\ 
$\mathcal{A}(k,\hat{\bm{k}}\cdot\hat{\lambda})$ & Power spectrum of a mode & Eq. \eqref{fouriercorr}\\
$\cos\phi$ & 2-D angle between sky-projected $\bm{r}$ and sky-projected $\hat{\lambda}$  & Eq. \eqref{eq:phifirst} \\
$\cos\gamma$ & Angle between line of sight and symmetry axis & Eq. \eqref{alfven2d}\\
$\mathcal{D}_n(R)$ & Multipole moment of centroid structure function & Eq. \eqref{centroidexp}\\
 \hline
\end{tabular}
\label{tab:params}
\end{table*}

\subsection{Centroids for optically thin emission lines}\label{sec:centroid}
In this section, we first review unnormalised centroids in the case when self-absorption is negligible. In this case, Eq. \eqref{eq8} gives
\begin{equation}
I_v(\bm{X})=\epsilon\rho_s(\bm{X},v)~,
\end{equation}
and, therefore Eq. \eqref{eq:unnormalised} gives
\begin{equation}\label{eq:unnormmaindef}
C(\bm{X})=\epsilon\int \mathop{\mathrm{d}v}v\rho_s(\bm{X},v)~.
\end{equation}
The usual approach in the study of centroids is to work in PPP space rather than PPV space. This can be achieved by writing Eq. \eqref{eq:unnormmaindef} as
\begin{align}\label{firstdef}
C(\bm{X})&=\epsilon\int \mathop{\mathrm{d}v}v\int\mathop{\mathrm{d}z}\rho(\bm{x})\Phi(v-u(\bm{x}))\nonumber\\
&=\epsilon\int\mathop{\mathrm{d}z}\rho(\bm{x})\int \mathop{\mathrm{d}v}v\Phi(v-u(\bm{x}))\nonumber\\
&=\epsilon\int\mathop{\mathrm{d}z}u(\bm{x})\rho(\bm{x})~,
\end{align}
where $\rho(\bm{x})$ is the real space density and $u(\bm{r})$ is the $z$-component of the turbulent velocity. 

However, in order to make a smooth connection between the optically thin case to the optically thick case, we derive centroids correlation function by working in the PPV space. This is straightforwardly achieved by utilising the theory of fluctuations of PPV space density $\rho_s(\bm{X})$ developed in \citetalias{lazarian2000velocity} and \citetalias{lazarian2004velocity}. Using Eq. \eqref{eq:unnormalised}, the correlation of centroids can be written as
\begin{align}\label{eq:firstcorr}
\xi(\bm{R})=\int_{-S}^S&\mathop{\mathrm{d}z_1}\int_{-S}^S\mathop{\mathrm{d}z_2}\xi_\rho(\bm{r})\int_{-\infty}^\infty\mathop{\mathrm{d}v_1}v_1\nonumber\\
&\int_{-\infty}^\infty\mathop{\mathrm{d}v_2}v_2\langle\Phi(v_1-u(\bm{x}_1))\Phi(v_2-u(\bm{x}_2))\rangle~,
\end{align}
where $\bm{R}=\bm{X}_1-\bm{X}_2$, $\bm{r}\equiv (\bm{R},z)=\bm{x}_1-\bm{x}_2$ and $\xi_\rho(\bm{r})$ is the density correlation function. In Eq. \eqref{eq:firstcorr}, $\langle..\rangle$ denotes the averaging over turbulent velocity $u(\bm{x})$ as a random quantity. The main assumption in writing Eq. \eqref{eq:firstcorr} is that the density and velocity fields are uncorrelated. This assumption has been tested in the past papers (eg. \citealt{esquivel2007statistics}), and seems to be sufficiently accurate for subsonic turbulence with density dispersion less than the mean density of a turbulent cloud. Assuming that the turbulent velocity is a Gaussian random vector, whose variance of the pairwise difference between two vectors is given by the structure function
\begin{equation}
D_{ij}(\bm{x}_2-\bm{x}_1)=\left\langle\left(u_i(\bm{x}_1)-u_i(\bm{x}_2)\right)\left(u_j(\bm{x}_1)-u_j(\bm{x}_2)\right)\right\rangle~,
\end{equation}
and the $z$-projected structure function is given by
\begin{equation}
D_z(\bm{x}_2-\bm{x}_1)=D_{ij}(\bm{x}_2-\bm{x}_1)\hat{z}_i\hat{z}_j~.
\end{equation}
With this, Eq. \eqref{eq:firstcorr} reduces to 
\begin{align}\label{centroidcorrm}
\xi(\bm{R})=\frac{1}{2\upi}\int_{-S}^S\mathop{\mathrm{d}z}\left(1-\frac{|z|}{2S}\right)\int_{-\infty}^\infty\mathop{\mathrm{d}v}\int_{-\infty}^\infty\mathop{\mathrm{d}v_+}\left(v_+^2-\frac{v^2}{4}\right)\nonumber\\
\frac{\xi_\rho(\bm{r})}{{\sqrt{D_z(\bm{r})+2\beta_{\text{T}}}}}
\exp\left[-\frac{v^2}{2(D_z(\bm{r})+2\beta_{\text{T}})}\right]\sqrt{\frac{2}{D^+(S, \bm{r})}}\nonumber\\
\exp\left[-\frac{v_+^2}{D^+(S, \bm{r})}\right]~,
\end{align}
where
\begin{equation}
D^+(S, \bm{r})\equiv \beta_{\text{T}}+D_z(S)-D_z(\bm{r})/2~,
\end{equation}
and $D_z(\bm{r})$ is the $z$-projected velocity structure function, $\beta_{\text{T}}\equiv k_BT/m$ is the thermal broadening, $m$ being the mass of atoms, $T$ being the temperature and $k_B$ being the Boltzmann constant. After performing the integration over $v$, we finally obtain
\begin{equation}\label{maincorr}
\xi(\bm{R})=\frac{1}{2}\int_{-S}^S\mathop{\mathrm{d}z}\left(1-\frac{|z|}{2S}\right)\xi_\rho(\bm{r})(D_z(S)-D_z(\bm{r}))~.
\end{equation}
Notice that our formalism cleanly shows how thermal effects drop out in centroids upon carrying out the integral in Eq. \eqref{centroidcorrm} to obtain Eq. \eqref{maincorr}. This shows that turbulence velocity spectrum can be recovered with centroids regardless of the temperature\footnote{For very hot plasmas, noise levels can distort centroid statistics. See Sec. \ref{finiteres} for more clarification.}, which is distinct from other techniques (e.g. VCA).

The centroids structure function is defined as
\begin{equation}\label{eq:defcenstr}
\mathcal{D}(R)=\left\langle\left[C(\bm{X}_1+\bm{R})-C(\bm{X}_1)\right]^2\right\rangle~.
\end{equation}
Utilising Eqs. \eqref{maincorr} and \eqref{eq:defcenstr}, we finally obtain the centroid structure function
\begin{align}\label{correlation}
\mathcal{D}(\bm{R})\approx\int_{-S}^S\mathop{\mathrm{d}z}\bigg\{D_z(S)\left(\xi_\rho(0,z)-\xi_\rho(\bm{r})\right)+[\xi_\rho(\bm{r})D_z(\bm{r})\nonumber\\
-\xi_\rho(0,z)D_z(0,z)]\bigg\}~.
\end{align}

With the assumption of zero correlation between density and velocity, the above result for optically thin line is identical to that obtained in LE03, where the same result was obtained by directly utilising Eq. \eqref{firstdef}. Working from first principles in the PPV space, as is done in this paper, is especially useful to deal with centroids in the presence of self-absorption. 

For a constant density field and at $R\ll S$, the centroid structure function is
\begin{equation}
\mathcal{D}(R)\propto R^{1+\nu}~,
\end{equation}
which is the regular centroid scaling. We use this scaling further in this paper.

\section{Centroids for optically thick emission lines}\label{othick}
With the introduction of centroids in PPV space, the extension of centroids to an absorbing media is straightforward. In this case, using full expression for intensity given by Eq. \eqref{eq8} in Eq. \eqref{eq:unnormalised} yields
\begin{equation}\label{eq:abcendef}
C(\bm{R})=\frac{\epsilon}{\alpha}\int \mathop{\mathrm{d}v}v\left[1-\mathrm{e}^{-\alpha\rho_s(\bm{X},v)}\right]~.
\end{equation}
Using Eqs. \eqref{eq:defcenstr} and \eqref{eq:abcendef} and following \citetalias{lazarian2004velocity}, one can obtain
\begin{align}
&\mathcal{D}(R)=\frac{\epsilon^2}{\alpha^2}\int_{-\infty}^\infty\mathop{\mathrm{d}v_1}v_1\int_{-\infty}^\infty\mathop{\mathrm{d}v_2}v_2
\bigg\langle \mathrm{e}^{-\alpha(\rho_{11}+\rho_{12})}\Big(1+\nonumber\\
&\mathrm{e}^{-\alpha(\rho_{22}+\rho_{21}-\rho_{11}-\rho_{12})}
-\mathrm{e}^{-\alpha(\rho_{21}-\rho_{11})}-\mathrm{e}^{-\alpha(\rho_{22}-\rho_{12})}\Big)\bigg\rangle~,
\end{align}
where $\rho_{ij}\equiv\rho_s(\bm{X}_i, v_j)$. Note that the exponential terms inside the parenthesis are split in such a way that the velocity is the same for two terms that make up a difference. As explained in \citetalias{lazarian2004velocity}, the reason for this arrangement is that at small separations $R$, the these terms inside are small and therefore the exponential can be expanded retaining only the leading order terms. This leads to
\begin{align}
\mathcal{D}\approx \epsilon^2\int_{-\infty}^\infty\mathop{\mathrm{d}v_1}v_1\int_{-\infty}^\infty\mathop{\mathrm{d}v_2}v_2
\big\langle \mathrm{e}^{-\alpha(\rho_{11}+\rho_{12})}\nonumber\\
\left[\left(\rho_{11}-\rho_{21}\right)\left(\rho_{12}-\rho_{22}\right)\right]\big\rangle~.
\end{align}
The above expression is the main expression that we study further. If one assumes density and velocity to be uncorrelated, this expression can be sufficiently simplified. The details of this simplification is presented in \citetalias{lazarian2004velocity}. Here we present the final result
\begin{align}\label{dabsor}
\mathcal{D}(\bm{R})\propto\epsilon^2\int_0^S\mathrm{d}z\int_{-\infty}^\infty\mathop{\mathrm{d}v}\mathrm{e}^{-\frac{\alpha^2}{2}\tilde{d}_s(0,v)}[\mathcal{W}(\bm{R},z,v)d_s(\bm{R},z, v)\nonumber\\
-\mathcal{W}(0,z,v)d_s(0,z,v)]~,
\end{align}
where
\begin{align}\label{windowself}
\mathcal{W}(\bm{R},z,v)=\int_{-\infty}^\infty\mathop{\mathrm{d}v_+}\left(v_+^2-\frac{v^2}{4}\right)\sqrt{\frac{2}{D^+(S, \bm{r})}}\nonumber\\
\exp\left[-\frac{v_+^2}{D^+(S, \bm{r})}\right]~,
\end{align}
and
\begin{equation}
d_s(\bm{R},z,v)=\frac{1}{\sqrt{D_z(\bm{r})+2\beta_{\text{T}}}}\exp\left[-\frac{v^2}{2(D_z(\bm{r})+2\beta_{\text{T}})}\right]~.
\end{equation}
In Eq. \eqref{dabsor}, $d_s(\bm{R}, v)=d_v(\bm{R}, v)+d_{\rho}(\bm{R}, v)$. We focus on the case when velocity term is dominant.
Our treatment to obtain Eq. \eqref{dabsor} is valid only when  $\alpha^2(d_s(R,v)-d_s(0,v)<1$ so that non-linear effects are negligible. A simpler, but loose, condition for velocity term is $\alpha^2d_v(R,0)<1$. This condition can be formally written by noting that the asymptote of $d_v(R,0)$ is given by $d_v(R,0)\sim R^{1-\nu/2}$ where we have omitted numerical pre-factors that are of order of unity. Restoring proper dimensionality, one can estimate the scales for which our treatment is applicable and this is given by
\begin{equation}\label{scalecon}
\alpha^2\langle\rho_s\rangle^2<\left(\frac{S}{R}\right)^{1-\nu/2}~.
\end{equation}
If the above condition is not fulfilled, one will see non-linear behaviour of the centroid structure function instead of power law in $R$. Note that the expression for centroid structure function derived in this paper (c.f Eq. \ref{dabsor}) and intensity structure function derived in \citetalias{lazarian2004velocity} looks similar. Closer inspection shows that the difference comes through the factor $\mathcal{W}(v)$. It is also important to note that the $R$ dependence in the $\mathcal{W}(v)$ given by Eq. \eqref{windowself} is weak, as $\beta_{\text{T}}+D_z(S)-D_z(\bm{r})/2$ is nearly a constant. 
For strong self-absorption non-linear effects are important. In the case of moderate self-absorption Eq. \eqref{dabsor} can be written as
\begin{equation}\label{windowabs}
\mathcal{W}(\bm{R},z,v)= \sqrt{\frac{\upi}{2}}\left(\beta_{\text{T}}+D_z(S)-\frac{D_z(\bm{r})}{2}-\frac{v^2}{2}\right)~.
\end{equation}
The main effect of absorption is the introduction of an exponentially suppressing factor $\exp[-\alpha^2\tilde{d}_s(0,v)/2]$. This is effectively an window, which tells us how integration over $v$ should be carried out. The critical width of absorption window is given by
\begin{equation}\label{window}
\alpha^2d_s(0,v_{\text{ab}})= 1~.
\end{equation}
The effects of absorption become strong for $v>v_{\text{ab}}$. Note that for weak self-absorption, i.e. for small $\alpha$, $v_{\text{ab}}$ is large and vice-versa.
Taking into consideration the power law behaviour of velocity structure function and density correlation function, one can solve for asymptotes of $d_s(0,v)$ to obtain (see \citetalias{lazarian2004velocity} for details)
\begin{align}\label{criticalv}
v_{\text{ab}}&\approx\sqrt{D_z(S)+2\beta_{\text{T}}}(\alpha\bar{\rho})^{\frac{2\nu}{\nu-2(1-\nu_\rho)}},\qquad \nu>(2/3)(1-\nu_\rho)\\
\label{criticalv1}
v_{\text{ab}}&\approx\sqrt{D_z(S)+2\beta_{\text{T}}}(\alpha\bar{\rho})^{-1},\qquad \qquad \text{ }\text{ }\text{ } \nu<(2/3)(1-\nu_\rho)~,
\end{align}
where the mean density $\bar{\rho}=\langle\rho_s\rangle (D_z(S)+2\beta_{\text{T}})^{1/2}/S$. Eqs. \eqref{criticalv} and \eqref{criticalv1} are valid only when  absorption is moderate so that $\alpha\bar{\rho}<1$.

At this point, we ask the following question: in the presence of absorption, can the velocity spectra be recovered? The subsequent analysis in this section is focused to answer this question. Before we start, we would first like to make some remark on the VCA technique developed in \citetalias{lazarian2000velocity} and \citetalias{lazarian2004velocity}, because some of the asymptotes of the centroid structure function in the presence of self-absorption will be similar to that in VCA. The VCA was developed for intensity (unlike centroids where we multiply LOS velocity and intensity). In VCA, one can control thickness of velocity slice from which data is analysed (see \citetalias{lazarian2000velocity} and \citetalias{lazarian2004velocity}). In a thin slice regime, the velocity window is essentially a delta function, and one can see effects turbulence velocity on intensity in this regime. In fact, a small scale asymptote of pure velocity contribution to the intensity structure function follows the scaling $R^{1-\nu/2}$, which is the thin-slice asymptote. On the other hand, for thick slice in an optically thin media, one essentially integrates over the entire range of velocity with a flat function, and in this regime velocity effects are erased and only density effects are manifested. It is important to note that to produce `thin-slice' asymptote, the velocity cut-off introduced by the window has to be less than $\sqrt{D_z(\bm{R})}$.

The whole point of the above review of the VCA is the following: although centroids are for integrated lines (and thus notion of `slice thickness' does not exist), self-absorption introduces an `effective slice thickness'. In fact, when self-absorption is large, both centroids and VCA yield the same `thin-slice' result. This can be seen from Eq. \eqref{centroidcorrm}, and noting the fact that $\beta_{\text{T}}+D_z(S)-D_z(\bm{r})/2$ is nearly a constant. In the case when self-absorption is negligible, centroids are defined effectively in `thick-slice' limit (as there is no window in the definition of centroids), and therefore this `thin-slice' asymptote can never be realised. The situation is different in the presence of self-absorption, as the factor $\exp[-\alpha^2d_s(0,v)/2]$ in Eq. \eqref{dabsor} introduced by self-absorption effectively acts like a window. Therefore, it is natural to expect that there exists a regime where even centroids produce `thin-slice' regime, which was relevant in the VCA. 

\begin{figure*}
\centering
\includegraphics[scale=0.5]{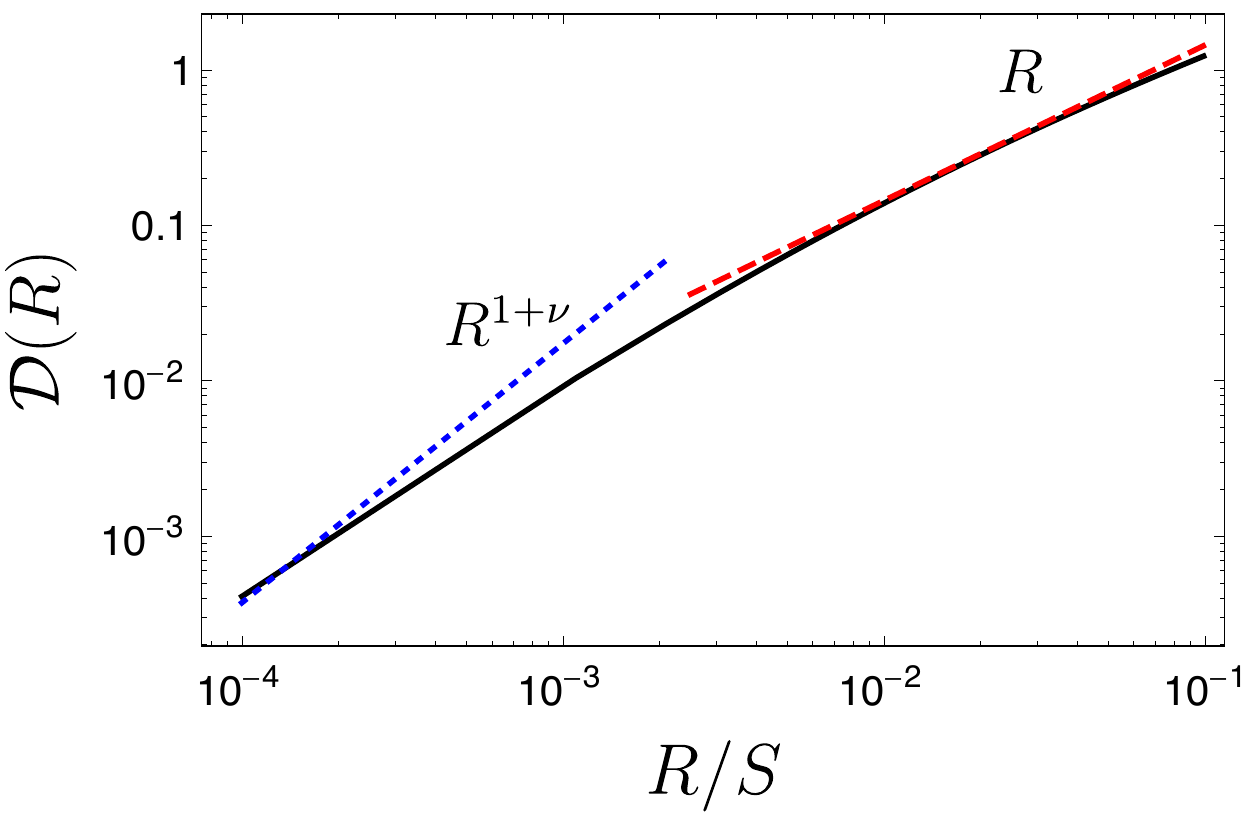}\hspace*{0.2cm}
\includegraphics[scale=0.5]{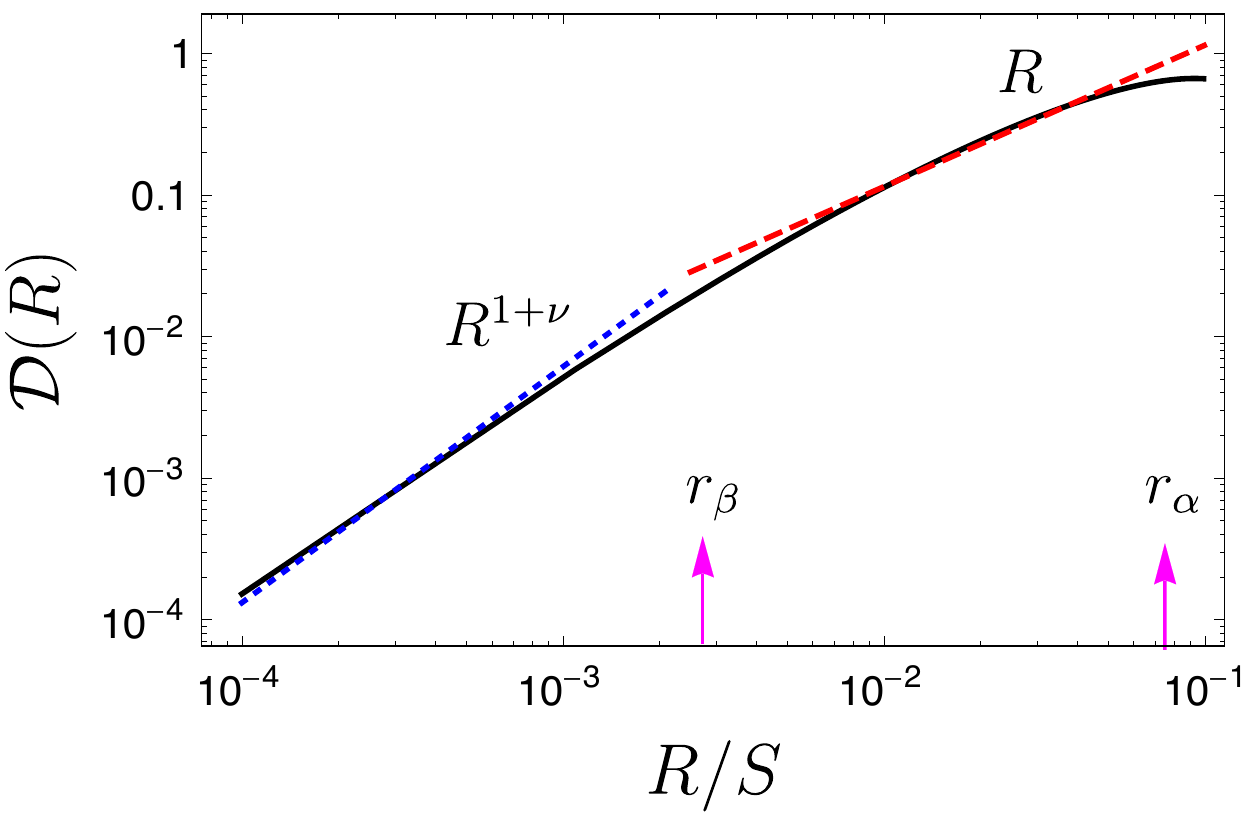}
\caption{Centroid structure function for different level of self-absorption for velocity spectrum with Kolomogorov index $\nu=2/3$ and for constant density field. The left-hand panel is for $\beta_{\text{T}}=0$, so that $r_{\beta}\equiv(\beta_{\text{T}}/D_z(S))^{1/\nu}=0$, and at $r_{\alpha}\equiv (v_c^2/(D_z(S)+2\beta_{\text{T}}))^{1/\nu}=0.2$. Right-hand panel is for different parameters defining thermal broadening and the level of self-absorption $r_\beta$ and $r_\alpha$, that are shown clearly in the figure. These panels clearly show two asymptotes scaling described in the text.}
\label{fig1}
\end{figure*}

In the case when $\nu\geq 2/3$, one can combine Eqs. \eqref{criticalv} and \eqref{scalecon} to obtain the condition for the validity of linear expansion $v_{\text{ab}}^2\geq (D_z(S)+2\beta_{\text{T}})(R/S)^{\nu}$, meaning that the critical velocity cut-off be larger than the rms velocity $D_z(S)$ of turbulent field. The implication is clear: in the case $\nu\geq 2/3$ one cannot obtain the `thin-slice' asymptote $R^{1-\nu/2}$. For $\nu<2/3$ one might be able to see power law behaviour $R^{1-\nu/2}$ before non-linear behaviour appears. On the other hand, at some intermediate scale $R$, absorption sets the width of the velocity kernel, and as explained in \citetalias{lazarian2004velocity} is of the order of $\Delta V\sim R^{\nu/2}$. In this situation, the scaling of the centroids is $R^{1-\nu/2}\Delta V\sim R$, and like in the VCA, centroids show an intermediate universal regime as well. In this regime, centroids lose information on the spectral slope of the velocity field. If one considers even smaller scale, then one should be able to recover the usual centroid scaling. In fact, if the velocity cut-off $v_{\text{ab}}$ introduced by self-absorption is much larger the velocity dispersion of the turbulent field, one should be able to recover the usual centroid scaling $R^{1+\nu}$. In the sub-sonic regime, where $D_z(R)\ll \beta_{\text{T}}$, one should again be able to recover the usual centroid scaling. This is one of the major advantage of centroids over VCA, which loses information about velocity spectrum at sub-sonic scales. These scaling are summarised in Table \ref{tab:absorpv}, and part of the asymptotes are shown in Fig.  \ref{fig1}.

In the absence of absorption, centroids are obtained by integrating over the entire line of sight velocity, and  therefore for a constant density field, the asymptote of centroid structure function scales as $\mathcal{D}(R)\sim R^{1+\nu}$. In the presence of absorption, although one should formally integrate over the entire line of sight velocity, the integration range is effectively set by the extent of absorption. In this case, depending on the extent of absorption one may or may not be able to obtain the same scaling as in the absence of absorption. In fact, as shown in Fig. \ref{fig1} and as summarised in Table \ref{tab:absorpv}, the asymptote might show different scaling at different lags $R$.

\begin{table*}
\centering
\caption{Scaling of centroid structure function arising from pure velocity effects in the presence of absorption. The scaling $R^{1-\nu/2}$ is only present for $\nu<2/3$. The corresponding scaling in the case of VCA is also presented for comparison.}
\begin{tabular}{l c c c}  
\hline
\emph{Scale range} & \emph{Centroid scaling} & \emph{VCA scaling (integrated lines)} & \emph{Regime} \\
\hline
$R/S<\left(\frac{\beta_{\text{T}}}{D_z(S)}\right)^{1/\nu}$ & $\mathcal{D}(R)\propto R^{1+\nu}$ & Velocity effects erased & Subsonic\\

$R/S\ll \left(\frac{v_{\text{ab}}^2}{D_z(S)+2\beta_{\text{T}}}\right)^{1/\nu}$ & $\mathcal{D}(R)\propto R^{1+\nu}$ & $\mathcal{D}(R)\propto R^{1+\nu/2\dagger}$ & Thick slice\\

$R/S<\left(\frac{v_{\text{ab}}^2}{D_z(S)+2\beta_{\text{T}}}\right)^{1/\nu}$ & $\mathcal{D}(R)\propto R$ & $\mathcal{D}(R)\propto R$ & Intermediate\\

$\left(\frac{v_{\text{ab}}^2}{D_z(S)+2\beta_{\text{T}}}\right)^{1/\nu}<R/S<\left(\frac{v_{\text{ab}}^2}{D_z(S)+2\beta_{\text{T}}}\right)^{2/(2-\nu)}$ & $\mathcal{D}(R)\propto R^{1-\nu/2}$ & $\mathcal{D}(R)\propto R^{1-\nu/2}$ & Effectively$^{\dagger\dagger}$  thin slice\\

$R/S>\left(\frac{v_{\text{ab}}^2}{D_z(S)+2\beta_{\text{T}}}\right)^{2/(2-\nu)}$ & Not a power law & Not a power law & Strong absorption\\
\hline
\end{tabular}
\begin{tabular}{l}
$\text{ }^\dagger$ One will see saturation $R^2$ of the structure function after this scaling.\\
$\text{ }^{\dagger\dagger}$ The discussion on VCA involves integrated lines, and thus slice thickness is not set by  interferometers but `effectively' \\
$\text{ }\text{ }\text{ }\text{ }\text{ }$by the level of self-absorption.
\end{tabular}
\label{tab:absorpv}
\end{table*} 

The most important finding is that in the presence of absorption, `usual' centroids may not be recovered at all. In fact, as shown in Table \ref{tab:absorpv}, one might be able to see the usual scaling of $R^{1+\nu}$ only in a very restricted range of scales. Another important finding is that for sufficiently hot turbulent plasma, the centroids work well even when self-absorption is strong. This can be understood the following way: if thermal effects are strong, then the effect of self-absorption is diminished. This is because for high temperature $\rho_s$ becomes small, and therefore, the intensity obtained from Eq. \eqref{eq8} looks more like that for an optically thin case. Our result that centroids work well for high temperatures even in the presence of self-absorption is very distinct and useful result in comparison to VCA, which works well for recovering velocity statistics only when the velocity dispersion is larger than the thermal broadening.

\section{Centroids for absorption lines}\label{abline}
In the previous section (and also in the past works), emission lines were used to obtain centroids. However, if the turbulence cloud is between an observer and an extended emission source, one can also measure velocity centroids for absorption lines. The 
turbulent motions affect the line profile, and thus one should be able to study turbulence using absorption lines. An advantage of studying turbulence with absorption lines is that multiple lines with various optical depths can be used simultaneously \citepalias{kandel2016extending}. Moreover, centroids are not sensitive to the gradients of wings of line profile, as multiplication by $v$ (which is odd) to even gradients in left and right wings, and upon integration washes away the effect of these gradients. In the context of VCS absorption lines were studied in \citetalias{lazarian2008studying}. It was shown that even in the presence of strong absorption line, one can obtain information on the turbulent spectra if one uses logarithm of intensity (i.e. optical depth) instead of intensity, and studies turbulence using the wings of the absorption line. The connection between the study carried out in \citetalias{lazarian2008studying} and centroids is simple: instead of using centroids for intensity, one needs to use optical depth to define centroids. However, because centroids are obtained by integrating over entire spectral line, in contrast to \citetalias{lazarian2008studying}, we focus on spatial correlations between lines of sight.  

The profile of an absorption line is given by
\begin{equation}\label{eq:opticaldepth}
I(\bm{X})=I_0\mathrm{e}^{-\tau(\bm{X},v_1)}~,
\end{equation}
where $\tau(\bm{X},v_1)$ is the optical depth. If intrinsic line broadening is ignored (see \citetalias{lazarian2008studying} for details), the optical depth is given by
\begin{equation}
\tau(\bm{X},v_1)=\alpha(\nu_0)\int_0^S\mathop{\mathrm{d}z}\rho(\bm{X},z)\Phi(v_1-u(z))~,
\end{equation}
The velocity centroid weighted by the optical depth can be defined as
\begin{equation}\label{eq:abdepth}
C(\bm{X})=\int \mathop{\mathrm{d}v_1}v_1\log\left(\frac{I(\bm{X},v_1)}{I_0}\right)=\int \mathop{\mathrm{d}v_1}v_1\tau(\bm{X},v_1)~.
\end{equation}
Notice that to obtain centroids, we used logarithm of intensity. Useful property of the centroids is that one does not need to precisely know the base-level $I_0$, since $v_1\log I_0$ is an odd function, and vanishes upon performing the integration over $v_1$.

If one has information on the optical depth throughout the whole line, it is clear from Eq.~ \eqref{eq:abdepth} that by using logarithm of intensity of an absorption line, one can obtain the centroid structure function of the same form as that of optically thin emission lines (cf. Eq. \eqref{centroidcorrm}), and thus one can obtain the same information about turbulence statistics.

However, real world observations have noise associated with them and therefore, realistically the profile of an absorption line is given by
\begin{equation}
I(\bm{X})=I_0\mathrm{e}^{-\tau(\bm{X},v_1)}+N~,
\end{equation}
where $N$ is the noise. Clearly for optical depths $\tau>-\log(N/I_0)$,  the central part of the line is saturated below the noise level and the useful information is restricted to the wings of the line, where $\tau<-\log(N/I_0)$. For such case, we suggest to construct restricted centroids by integrating $\int \mathop{\mathrm{d}v_1}v_1\tau(\bm{X},v_1)$ just over  the wings of the line, where optical depths can be accurately determined. To maintain the properties of centroids one should
integrate over a symmetric pair of intervals of width $\Delta$ centered
at $v_0$ and $-v_0$ for the left and right wing (line is assumed to be centered at $v=0$).

As an illustration, let us consider wings of the line to be selected by a  sharp window.  In this case, the restricted centroids are given by
\begin{equation}\label{sharpdef}
C(\bm{X})=\int_{-v_0-\Delta/2}^{-v_0+\Delta/2} \mathop{\mathrm{d}v_1}v_1\tau(\bm{X},v_1)+\int_{v_0-\Delta/2}^{v_0+\Delta/2} \mathop{\mathrm{d}v_1}v_1\tau(\bm{X},v_1)~,
\end{equation}
where the center of the wing $v_0$ is of the order of $\sqrt{D_z(S)+2\beta_\text{T}}$. Formally, when $\Delta$ approaches $2v_0$ two wings overlap and the whole line is available for analysis.

Using the usual approach (see Eqs. \eqref{eq:firstcorr} and \eqref{maincorr}), one obtain the centroid correlation as
\begin{align}\label{evenodd}
\xi(\bm{R})=2\left[\xi_{11}(\bm{R})+\xi_{12}(\bm{R})\right]~,
\end{align}
where $\xi_{11}(\bm{R})$ denotes correlations within the same wing, and is given by
\begin{align}\label{even}
\xi_{11}(\bm{R})\propto\int_{-S}^S\mathop{\mathrm{d}z}\int_{v_0-\Delta/2}^{v_0+\Delta/2} \mathop{\mathrm{d}v_1}v_1\int_{v_0-\Delta/2}^{v_0+\Delta/2} \mathop{\mathrm{d}v_2}v_2\nonumber\\
\frac{\xi_\rho(\bm{r})}{{\sqrt{D_z(\bm{r})+2\beta_{\text{T}}}}}
\exp\left[-\frac{(v_1-v_2)^2}{2(D_z(\bm{r})+2\beta_{\text{T}})}\right]\sqrt{\frac{2}{D^+(S, \bm{r})}}\nonumber\\
\exp\left[-\frac{(v_1+v_2)^2}{4D^+(S, \bm{r})}\right]~,
\end{align}
and  $\xi_{12}(\bm{R})$ denotes correlation between two wings, and is given by
\begin{align}\label{odd}
\xi_{12}(\bm{R})\propto\int_{-S}^S\mathop{\mathrm{d}z}\int_{v_0-\Delta/2}^{v_0+\Delta/2} \mathop{\mathrm{d}v_1}v_1\int_{-v_0-\Delta/2}^{-v_0+\Delta/2} \mathop{\mathrm{d}v_2}v_2\ldots~,
\end{align}
where $\ldots$ denotes that the integrand in Eq. \eqref{odd} is the same as that in Eq. \eqref{even}, but the integration range over $v_1$ and $v_2$ are non-overlapping.
\begin{table}
\centering
\caption{Scaling of centroid structure function arising from pure velocity effects in absorption line study.}
\begin{tabular}{l c}  
\hline  
\emph{Scale range} & \emph{Centroid scaling} \\
\hline
$R/S\lesssim \left(\frac{2(\Delta^2+\beta_{\text{T}})}{D_z(S)}\right)^{1/\nu}$ & $\mathcal{D}(R)\propto R^{1+\nu}$\\
$R/S\gtrsim\left(\frac{2(\Delta^2+\beta_{\text{T}})}{D_z(S)}\right)^{1/\nu}$ & $\mathcal{D}(R)\propto R^{1-\nu/2}$\\
\hline
\end{tabular}
\label{tab:absorpline}
\end{table}

We now investigate two extreme limits: firstly, $\Delta\rightarrow 0$. In this case, the integration over a narrow range of $\Delta$ can be approximated by evaluating the integrand at central value of the integration range. Thus, we have
\begin{align}
\xi_{11}(\bm{R})\propto\Delta^2\int_{-S}^S\mathop{\mathrm{d}z}v_0^2
\frac{\xi_\rho(\bm{r})}{{\sqrt{D_z(\bm{r})+2\beta_{\text{T}}}}}
\sqrt{\frac{2}{D^+(S, \bm{r})}}\nonumber\\
\exp\left[-\frac{v_0^2}{4D^+(S, \bm{r})}\right]~,
\end{align}
and 
\begin{align}
\xi_{12}(\bm{R})\propto-\Delta^2\int_{-S}^S\mathop{\mathrm{d}z}v_0^2
\frac{\xi_\rho(\bm{r})}{{\sqrt{D_z(\bm{r})+2\beta_{\text{T}}}}}\sqrt{\frac{2}{D^+(S, \bm{r})}}\nonumber\\
\exp\left[-\frac{2v_0^2}{(D_z(\bm{r})+2\beta_{\text{T}})}\right]~.
\end{align} 
Since $v_0^2\sim D^{+}(S,\bm{r})$, it is clear that due to an exponential suppression $\xi_{12}(\bm{R})\ll \xi_{11}(\bm{R})$, and therefore for $\Delta\sim 0$ the correlation is given by
\begin{align}\label{dnarrow}
\xi(\bm{R})\approx 2\xi_{11}(\bm{R})\propto\Delta^2\int_{-S}^S\mathop{\mathrm{d}z}v_0^2
\frac{\xi_\rho(\bm{r})}{{\sqrt{D_z(\bm{r})+2\beta_{\text{T}}}}}~.
\end{align}
At small scales, the factor $D^+(\bm{r})$ is close to constant, and therefore the asymptote in Eq. \eqref{dnarrow} fully depends on the form of the integrand $\xi_\rho(\bm{r})/\sqrt{D_z(\bm{r})+2\beta_{\text{T}}}$.  Looking at Eq. \eqref{dnarrow}, one can see that  for $D_z(\bm{R})>2\beta_{\text{T}}$, one obtains an asymptote $R^{1-\nu/2}$ in the case of constant density field. This formally corresponds to the `thin-slice' regime in the context of VCA. Indeed, as $\Delta$ and $\beta_{\text{T}}$ become smaller, the integration channel slice become thinner, and this corresponds to returning to the thin slice regime of VCA. In the case when $\Delta$ is finite, $D_z(\bm{R})>2(\beta_{\text{T}}+\Delta^2)$ needs to be satisfied in order to achieve thin-slice asymptote.

\begin{figure}
\centering
\includegraphics[scale=0.5]{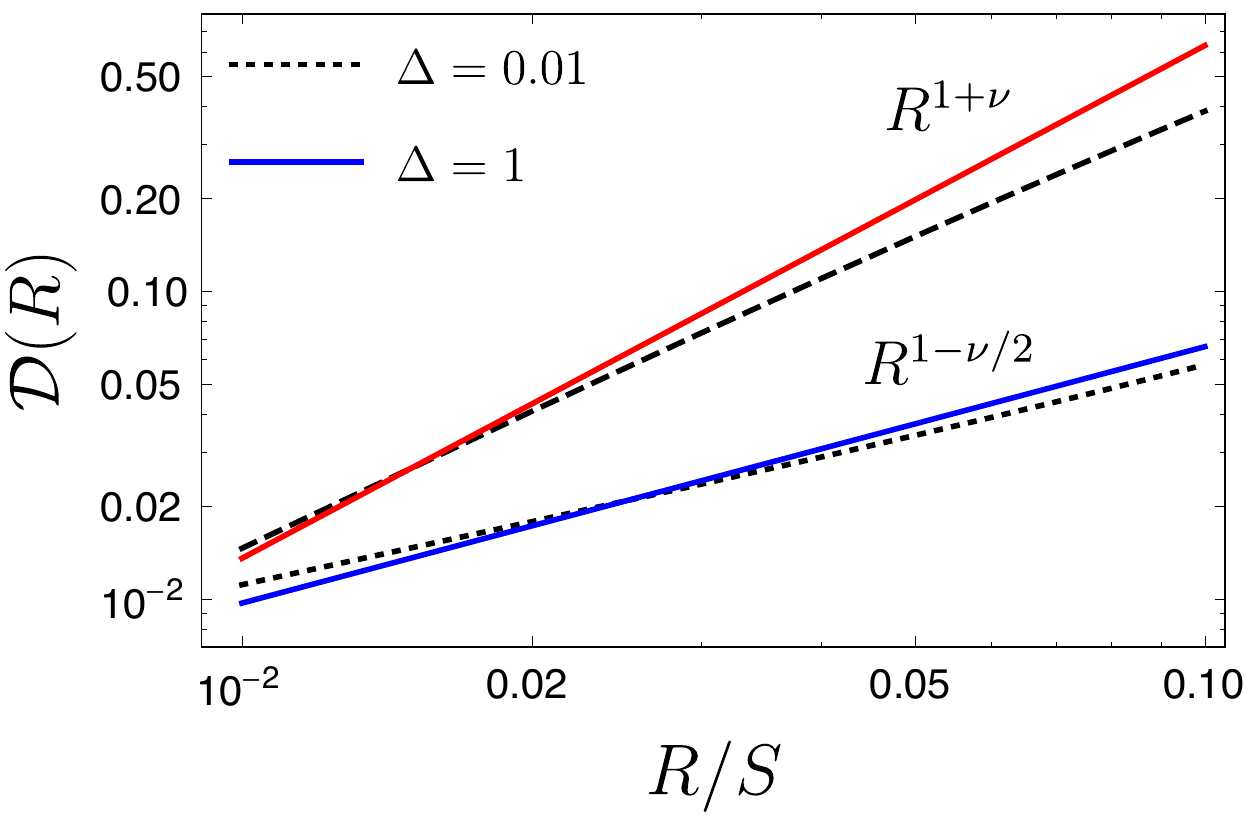}
\caption{Centroid structure for broadening parameter $\Delta=0.01$ and   $1$, and for a constant density field. For small $\Delta$, the asymptote $R^{1-\nu/2}$ is reached fast, while for large $\Delta$ the asymptote $R^{1+\nu}$ is extended over large range of $R$.The effect of increasing temperature is {\it effectively} the same as increasing $\Delta$. Example curves are produced for velocity field with Kolmogorov index $\nu=2/3$.}
\label{fig2}
\end{figure}

The second case of interest is $\Delta= 2v_0$, which formally corresponds to two wings touching each other, which selects central section of the line of width $2\Delta$. Using Eq. \eqref{sharpdef}, it can be shown that
\begin{align}\label{dwide}
\xi&(\bm{R})\propto\int_{-S}^S\mathop{\mathrm{d}z}\int_{-\Delta}^{\Delta} \mathop{\mathrm{d}v_1}v_1\int_{-\Delta}^{\Delta} \mathop{\mathrm{d}v_2}v_2
\frac{\xi_\rho(\bm{r})}{{\sqrt{D_z(\bm{r})+2\beta_{\text{T}}}}}\nonumber\\
&\exp\left[-\frac{(v_1-v_2)^2}{2(D_z(\bm{r})+2\beta_{\text{T}})}\right]\sqrt{\frac{2}{D^+(S, \bm{r})}}
\exp\left[-\frac{(v_1+v_2)^2}{4D^+(S, \bm{r})}\right]~.
\end{align}
It is easy to see from Eq. \eqref{dwide} that when $\Delta\rightarrow \infty$, the usual centroids are recovered. A less stringent criteria to recover centroids when $\Delta\sim \sqrt{D_z(S)+2\beta_\text{T}}$ is to study turbulence at small lags $R$ satisfying $D_z(R)<2(\Delta^2+\beta_{\text{T}})$. This condition can be achieved either for sufficiently short scales or for sufficiently large $\Delta$ and $\beta_{\text{T}}$. Physically, large $\Delta$ means entire line is available for centroids statistics, and it is natural to expect to recover usual centroids in such limit. Numerical integration confirms this, and this has been clearly shown in Fig. \ref{fig2} and Table \ref{tab:absorpline}, where it is shown that with at small scales $R$, the velocity spectrum is correctly given by the centroids. Notice that with the thermal effects included, there is larger range of lag $R$, which yields the usual centroid correlation.

\section{Centroid anisotropy: general formalism}\label{sec:genanisotropy}
In this section, the study of centroids will be carried out keeping in mind that the ISM is magnetized and therefore there exists a preferred direction, which in the global frame of reference is the direction of the mean magnetic field. Due to the presence of this preferred direction, turbulence is anisotropic; to be more precise, axisymmetric. This anisotropy is built in the general tensors representing velocity, density and magnetic field correlations. The study of magnetic correlation was carried out in \citetalias{lazarian2012statistical} , and the study of velocity correlation and its application to study intensity anisotropies was carried out in \citetalias{kandel2016extending}. In this paper, we present another way of studying anisotropies, which is through the study of anisotropy of centroids correlation. The main focus of this section is to develop a general formalism to study centroid anisotropy in the presence of {\it constant} density field. All the relevant notations used in this section (as well as other sections) are presented in Table \ref{tab:params}.

The Fourier component of a velocity field is in general given by
$\bm{v}(\bm{k})=a_{\bm{k}}\hat{\xi}(\hat{\bm{k}},\hat{\lambda}),$ where $\bm{k}$ is the wavevector, 
$a_{\bm{k}}$ is the random amplitude
of a mode (which in general has both real and imaginary parts) and $\hat{\xi}$ is the direction of allowed displacement in a plasma. With this definition, the velocity correlation in Fourier space is given by
\begin{align}\label{fouriercorr}
\langle v_i(\bm{k}) v_j^*(\bm{k}') \rangle
=\langle a_{\bm{k}}&a^*_{\bm{k}'}\rangle
\left(\hat{\xi}_{\bm{k}} \otimes \hat{\xi}^*_{\bm{k}'}\right)_{ij}\nonumber\\
&\equiv \mathcal{A}(k,\hat{\bm{k}}\cdot\hat{\lambda})\left(\hat{\xi}_{\bm{k}} \otimes \hat{\xi}^*_{\bm{k}}\right)_{ij}\delta(\bm{k}-\bm{k}')~,
\end{align}
where $\mathcal{A}(k,\hat{\bm{k}}\cdot\hat{\lambda})=\langle \hat{a}_{\bm{k}}\hat{a}^*_{\bm{k}}\rangle$ is the power spectrum, which for anisotropic turbulence  depends on the angle 
$\mu_{k}\equiv \hat{\bm{k}}\cdot\hat{\lambda}$.
To obtain velocity correlation tensor in the real space one needs to carry out Fourier transform of Eq. \eqref{fouriercorr}
\begin{equation}\label{eq:realspacevcorr}
\left\langle v_i(\bm{x}_1) v_j(\bm{x}_1+\bm{r})\right\rangle =\frac{1}{(2\upi)^3}\int \mathop{\mathrm{d}^3\bm{k}}\mathrm{e}^{\mathrm{i}\bm{k}\cdot\bm{r}}\mathcal{A}(k,\hat{\bm{k}}\cdot\hat{\lambda})\left(\hat{\xi}_{\bm{k}} \otimes \hat{\xi}^*_{\bm{k}}\right)_{ij}~.
\end{equation}
The centroid structure function in the case of constant density for an optically thin medium is given by (cf. Eq. \eqref{correlation})
\begin{equation}
\mathcal{D}(R)\propto\int\mathop{\mathrm{d}z}(D_z(R,z)-D_z(0,z))~.
\end{equation}
In order to evaluate the above integral, we need to evaluate integral of the type $\int\mathop{\mathrm{d}z}\langle v_iv_j\rangle $. This can effectively be obtained by noting  that the integration over the entire line of sight is equivalent to setting $k_z=0$ in the spectral domain \citepalias{lazarian2012statistical} . Therefore, we can write
\begin{equation}
\int\mathop{\mathrm{d}z}\langle v_iv_j\rangle=\frac{1}{(2\upi)^2}\int\mathop{\mathrm{d}^2\bm{K}}\mathrm{e}^{\mathrm{i}\bm{K}\cdot\bm{R}}\mathcal{A}(K,\hat{\bm{K}}\cdot\hat{\Lambda})\left(\hat{\xi}_{\bm{K}} \otimes \hat{\xi}^*_{\bm{K}}\right)_{ij}~.
\end{equation}
We use plane wave the plane wave expansion
\begin{equation}
\mathrm{e}^{\mathrm{i} \bm{K}\cdot\bm{R}}=\mathrm{e}^{\mathrm{i} KR\cos\zeta_R}=\sum_{n=-\infty}^\infty \mathrm{i}^nJ_n(K R)\mathrm{e}^{\mathrm{i}n\zeta_R}~,
\end{equation}
where $\cos\zeta_R=\hat{\bm{K}}\cdot\hat{\bm{R}}$.
Similarly, decomposing the two dimensional power spectra into series of harmonics
\begin{equation}\label{decompower}
\mathcal{A}(K,\hat{\bm{K}}\cdot\hat{\Lambda})=\sum_{p=-\infty}^\infty K^{-3-\nu}\hat{\mathcal{A}}_p\mathrm{e}^{\mathrm{i}p\zeta_{\Lambda}}~,
\end{equation}
where $\cos\zeta_{\Lambda}=\hat{\bm{K}}\cdot\hat{\bm{\Lambda}}$, we  obtain
\begin{align}\label{eq:phifirst}
\int\mathop{\mathrm{d}z}\langle v_iv_j\rangle=\frac{1}{(2\upi)^2}\int\mathop{\mathrm{d}\bm{K}}K^{-2-\nu}\sum_{n=-\infty}^\infty \mathrm{i}^n\mathrm{e}^{\mathrm{i}n\phi}J_n(K R)\nonumber\\
\sum_{p=-\infty}^\infty\hat{\mathcal{A}}_p\mathrm{e}^{-\mathrm{i}(n-p)\psi}\left(\hat{\xi}_{\bm{K}} \otimes \hat{\xi}^*_{\bm{K}}\right)_{ij}~,
\end{align}
where $\cos\psi=\hat{\bm{K}}\cdot\hat{\bm{\Lambda}}$. Due to the axisymmetric nature of the turbulence, only even $p$ is allowed in Eq. \eqref{decompower}. With this, we finally obtain the following form of centroid structure function
\begin{align}\label{maincentroidexpd}
\mathcal{D}(\bm{R})=\frac{1}{(2\upi)^2}\int\mathop{\mathrm{d}\bm{K}}K^{-2-\nu}\sum_{n=-\infty}^\infty \mathrm{i}^n\mathrm{e}^{\mathrm{i}n\phi}(J_n(0)\delta_{n0}\nonumber\\
-J_n(K R))\sum_{p=-\infty}^\infty\hat{\mathcal{A}}_p\mathrm{e}^{-\mathrm{i}(n-p)\psi}\left(\hat{\xi}_{\bm{K}} \otimes \hat{\xi}^*_{\bm{K}}\right)_{zz}~,
\end{align}
where $n$ is even due to the fact that $p$ is even. To study anisotropy of the structure function, it is convenient to  expand the structure function in series of two dimensional circular harmonics
\begin{equation}\label{centroidexp}
\mathcal{D}(R,\phi)=\sum_{n=-\infty}^{\infty}\mathcal{D}_n(R)\mathrm{e}^{\mathrm{i}n\phi}~,
\end{equation}
where $\mathcal{D}_n(R)$ is the multipole moment of the centroid structure function given by (cf. Eq. \eqref{maincentroidexpd})
\begin{align}\label{centriodcoeff}
\mathcal{D}_n(R)=C_n(\nu)\sum_{p=-\infty}^\infty\hat{\mathcal{A}}_p\mathcal{W}_{n-p}R^{1+\nu}~,
\end{align}
and  $\mathcal{W}_{p}$ is the spectral weight function, which is the integral of the tensor structure of a specific mode over the two dimensional angle $\psi$, given by
\begin{equation}
\mathcal{W}_{p}=\frac{1}{2\upi}\int_0^{2\upi}\mathop{\mathrm{d}\psi}\mathrm{e}^{-\mathrm{i}p\psi}\left(\hat{\xi}_{\bm{K}} \otimes \hat{\xi}^*_{\bm{K}}\right)_{zz}~,
\end{equation}
and 
\begin{align}
C_n(\nu)=\mathrm{i}^n\int\mathop{\mathrm{d}K}K^{-2-\nu}&(J_n(0)\delta_{n0}-J_n(K))\nonumber\\
&=-\frac{\mathrm{i}^n \Gamma \left[\frac{1}{2} (|n|-\nu -1)\right]}{2^{2+\nu}\Gamma \left[\frac{1}{2} (|n|+\nu +3)\right]}~.
\end{align}
Eqs. \eqref{centroidexp} and \eqref{centriodcoeff} are the main equations that will be used subsequently to obtain centroid structure function of each MHD modes. An useful parameter for comparison with past numerical work is the isotropy degree, defined as
\begin{equation}
\text{Isotropy Degree}=\frac{\mathcal{D}(R,\phi=0)}{\mathcal{D}(R,\phi=\upi/2)}~.
\end{equation}
\section{Centroids for different MHD modes}\label{sec:mhdmodes}
The properties of MHD turbulence depend on the degree of magnetization. The Alfv\'en Mach number $M_{\text{A}}=V_L/V_A$, where $V_L$ is the injection velocity at the scale $L$ and $V_A$ is the Alfv\'en velocity, presents a useful measure of magnetization. Depending on whether $M_{\text{A}}>1$, $M_{\text{A}}=1$ or $M_{\text{A}}<1$, turbulence can be super-Alfv\'enic, trans-Alfv\'enic or sub-Alfv\'enic. For $M_{\text{A}}\gg 1$ magnetic forces are not important at large scales and the cascade should be similar to ordinary hydrodynamic cascade in the vicinity of the injection scale. The seminal paper by \citet[hereafter \citetalias{goldreich1995toward}]{goldreich1995toward} ushered the modern understanding of MHD turbulence. The \citetalias{goldreich1995toward} was formulated for trans-Alfv\'enic turbulence, and the generalization of \citetalias{goldreich1995toward} for sub-Alfv\'enic and super-Alfv\'enic cases can be found in \citet{lazarian1999reconnection}.  The original \citetalias{goldreich1995toward} theory was elaborated in further studies, in particular the concept of local system of reference (\citealt{lazarian1999reconnection}; \citealt{cho2000anisotropy}; \citealt{maron2001simulations}; \citealt{cho2002simulations}) was introduced. According to this concept, turbulent motions should not be viewed in the system of reference of the mean magnetic field as all earlier theories of MHD turbulence attempted to do, but in the system of reference of magnetic field comparable with the size of the eddies. However, from the point of view of the observational studies of the turbulence in a volume when the only available statistics are those averaged along the line of sight, the measurements should be carried out in the system of mean magnetic field, rather than the local system of reference. Therefore, one has to describe Alfv\'enic turbulence in the global system of reference  (see the discussions in \citealt{cho2002compressible}; \citealt{esquivel2005velocity}; \citetalias{lazarian2012statistical}). This modifies the available statistics. For instance, in the local system of reference \citetalias{goldreich1995toward} predicts the existence of two different energy spectra, namely, the parallel and perpendicular, in the global system of reference  only the spectrum of dominant perpendicular fluctuations is available. Similarly, while in the local system of reference the anisotropy increases with the decrease of size of the eddies, the anisotropy stays constant in the global system of reference.

MHD turbulence can be presented as a superposition of interacting fundamental modes, i.e. Alfv\'en, slow and fast. The first theoretical considerations in favour of this were given in \citetalias{goldreich1995toward} (see also \citealt{lithwick2001compressible}), which were extended and numerically tested in \cite{cho2002compressible, cho2003compressible} and in \cite{kowal2010velocity}. Because the compressible and incompressible modes weakly exchange their energy\footnote{This was recently shown to be also true for relativistic MHD turbulence in Makoto \& Lazarian 2016.} (\citealt{cho2002compressible}), it is possible to consider the modes separately.

With this background discussion of MHD turbulence, we are ready to proceed to study centroids anisotropy in detail. In this section, we will employ our previous expression to obtain centroids for different MHD modes. We will use the tensor structures obtained by \citetalias{kandel2016extending} for our discussion. These expressions will be used in the next section to specific MHD mode in order to study anisotropy due to each MHD mode.

\subsection{Alfv\'en mode}
\begin{figure*}
\centering
\includegraphics[scale=0.4]{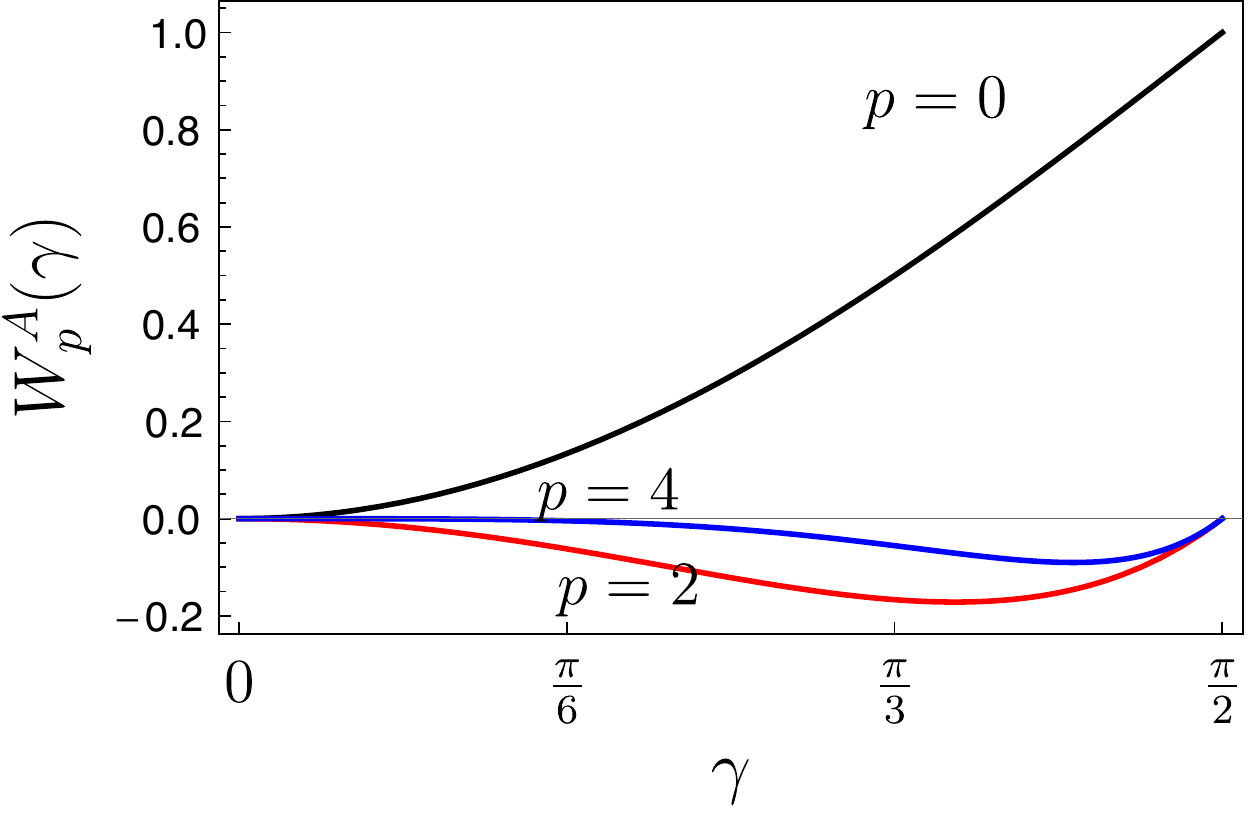}\hspace*{0.2cm}
\includegraphics[scale=0.4]{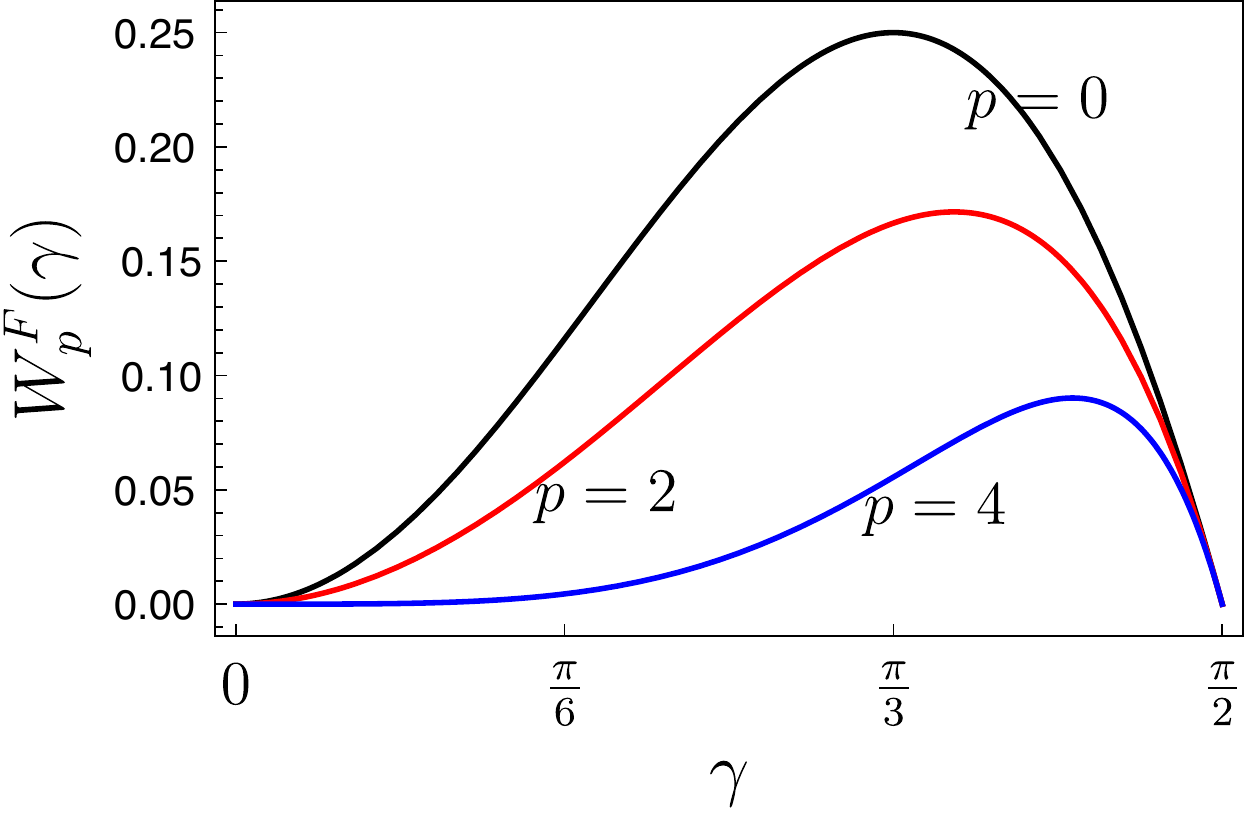}\hspace*{0.2cm}
\includegraphics[scale=0.4]{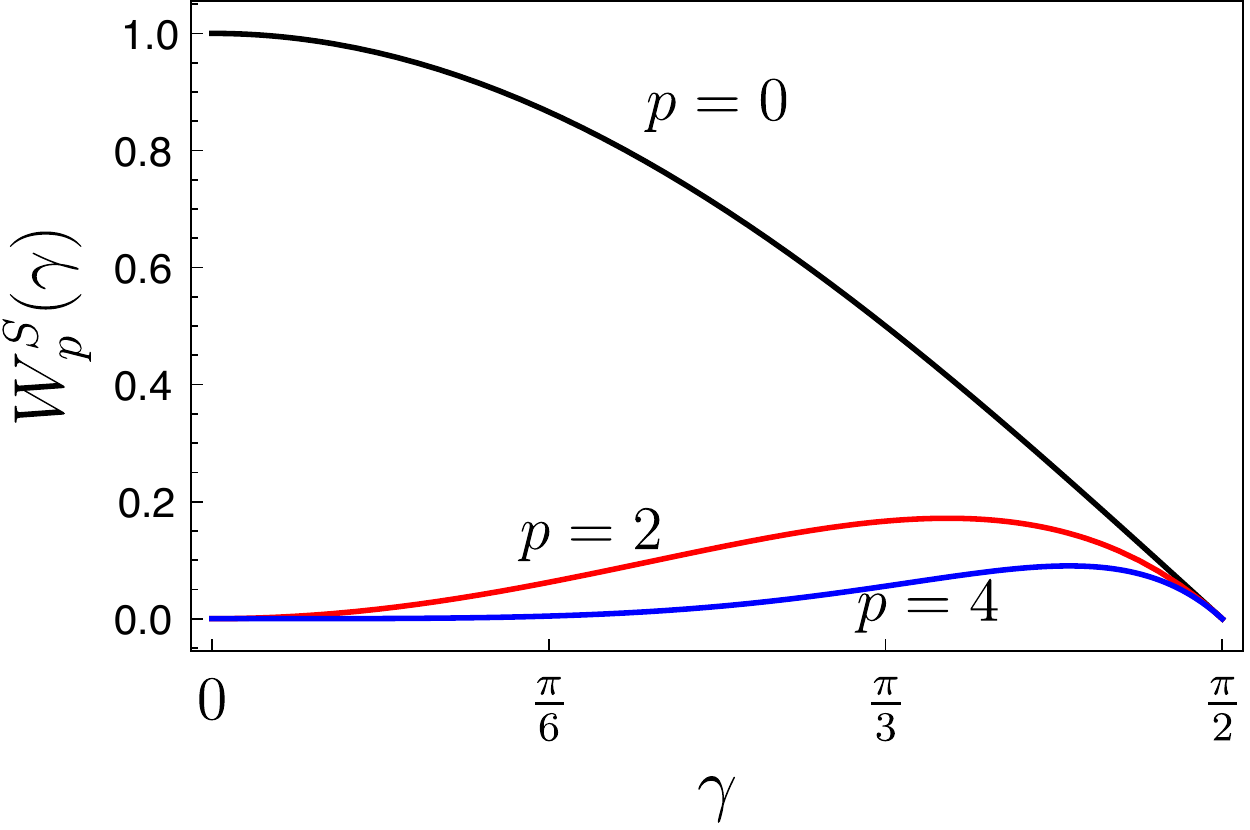}
\caption{From left-hand to right-hand: spectral function of Alfv\'en mode $W_p^A(\gamma)$ (left-hand), low-$\beta$ fast mode $W_p^F(\gamma)$ (center) and high-$\beta$ slow mode $W_p^S(\gamma)$ (right-hand) for various index $p$ (which is $n-p$ in Eq. \eqref{centriodcoeff}. }
\label{fig3}
\end{figure*}

Alfv\'en modes are incompressible, and their tensor should also reflect this. The correlation function of Alfv\'en mode is obtained by employing the condition that the displacement of this mode in a plasma is orthogonal to the wave-vector and the direction of magnetic field, and is given by (see \citetalias{kandel2016extending})
\begin{align}\label{eq:alfventensor}
&\left(\hat{\xi}_{\bm{k}} \otimes \hat{\xi}^*_{\bm{k}}\right)_{ij}\nonumber\\
&=\left(\delta_{ij}-\hat{k}_i\hat{k}_j\right)-\frac{(\hat{\bm{k}}\cdot\hat{\lambda})^2\hat{k}_i\hat{k}_j+\hat{\lambda}_i\hat{\lambda}_j-(\hat{k}\cdot\hat{\lambda})(\hat{k}_i\hat{\lambda}_j+\hat{k}_j\hat{\lambda}_i)}{1-(\hat{\bm{k}}\cdot\hat{\lambda})^2}~.
\end{align}
In \citetalias{lazarian2012statistical} , the first part of the above tensor was referred to as `$E$-type' while the second part was referred to as `$F$-type'. Both $E$ and $F$-type parts are divergence free, and therefore the velocity field in Alfv\'en modes is purely solenoidal. For an isotropic power spectra $\mathcal{A}$, the $E$-part yields isotropic correlation tensor, while $F$-part still gives rise to anisotropy.
Making use of Eq. \eqref{eq:alfventensor}, we obtain
\begin{equation}\label{alfven2d}
\left(\hat{\xi}_{\bm{K}} \otimes \hat{\xi}^*_{\bm{K}}\right)_{zz}=1-\frac{\hat{\lambda}_z\hat{\lambda}_z}{1-(\hat{\bm{K}}\cdot\hat{\Lambda})^2}=\sin^2\gamma\frac{\sin^2\psi}{1-\sin^2\gamma\cos^2\psi}~,
\end{equation}
where $\cos\gamma=\hat{r}\cdot\hat{\lambda}$, and $0\leq\gamma\leq\pi/2$. Note that correlation given by Eq. \eqref{alfven2d} vanishes at $\gamma=0$, which is expected as motions are perpendicular to the magnetic field. 
Making use of Eqs. \eqref{centriodcoeff} and \eqref{alfven2d}, the multipole moments of centroid structure function for Alfv\'en mode can be written as
\begin{equation}\label{alfvenstr}
\mathcal{D}_n(R)=C_n(2/3)\sum_{p=-\infty}^\infty\hat{\mathcal{A}}_p\mathcal{W}_{n-p}^AR^{5/3}~,
\end{equation} 
where $\hat{\mathcal{A}}_p$ is the coefficient of two dimensional harmonic expansion of power spectrum, and, as suggested in \citep{cho2002compressible}, is given by 
\begin{equation}
\hat{\mathcal{A}}_p=\frac{1}{2\upi}\int_{0}^{2\upi}\mathop{\mathrm{d}\psi}\mathrm{e}^{-\mathrm{i}p\psi}\exp\left[-M_{\text{A}}^{-4/3}\frac{|\cos\psi|\sin\gamma}{(1-\cos^2\psi\sin^2\gamma)^{2/3}}\right]~,
\end{equation}
and $\mathcal{W}_{n-p}^A$ spectral weight defined as
\begin{equation}\label{spectralintegral}
\mathcal{W}_{n-p}^A=\frac{1}{2\upi}\int_0^{2\upi}\mathop{\mathrm{d}\psi}\mathrm{e}^{-\mathrm{i}(n-p)\psi}\frac{\sin^2\gamma\sin^2\psi}{1-\sin^2\gamma\cos^2\psi}~.
\end{equation}
An analytical form of this spectral weight exists and is given by
\begin{equation}
\mathcal{W}_{n-p}^A=\delta_{p,n}-\cos\gamma\left(\frac{1-\cos\gamma}{\sin\gamma}\right)^{|n-p|}~.
\end{equation}
It is clear from Eqs. \eqref{alfvenstr} and \eqref{spectralintegral} that the centroid structure function of Alfv\'en mode vanishes at $\gamma=0$, which reflects that there is no LOS component of the 
Alfv\'en velocity when magnetic field is along the LOS. In the opposite case $\gamma=\upi/2$ when magnetic field is perpendicular to the LOS,   $\mathcal{W}_{n-p}^A=\delta_{pn}$, and multipole moments of the centroid structure function $\mathcal{D}_n(R)\propto \mathcal{A}_n$. It can be clearly seen from the left hand panel of Fig. \ref{fig3}. For general $\gamma$, it is also clear from the figure that the magnitude of the function $\mathcal{W}_{n-p}^A$ decays rapidly as $|n-p|$ increases. This means that for our practical purposes, it is enough to just few terms near $p\approx n$ in the sum presented in Eq. \eqref{alfvenstr}.

\begin{figure*}
\centering
\includegraphics[scale=0.4]{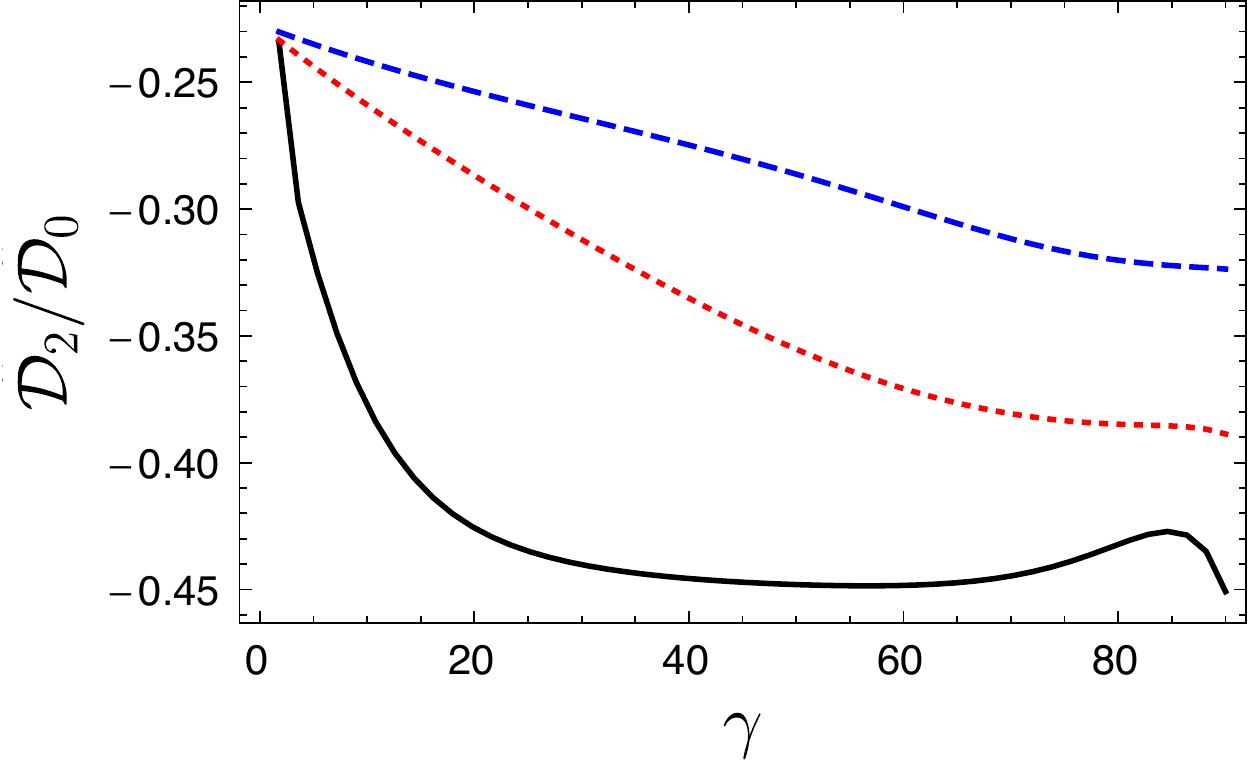}\hspace*{0.2cm}
\includegraphics[scale=0.4]{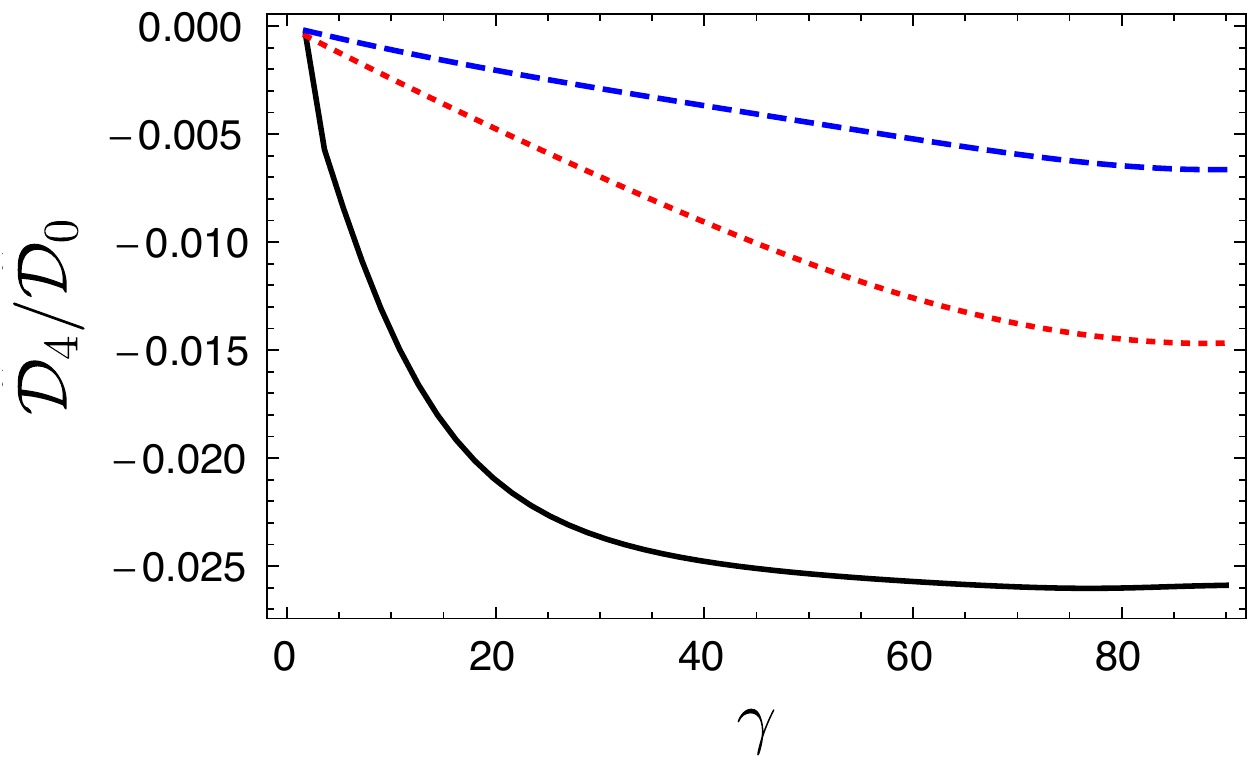}\hspace*{0.2cm}
\includegraphics[scale=0.4]{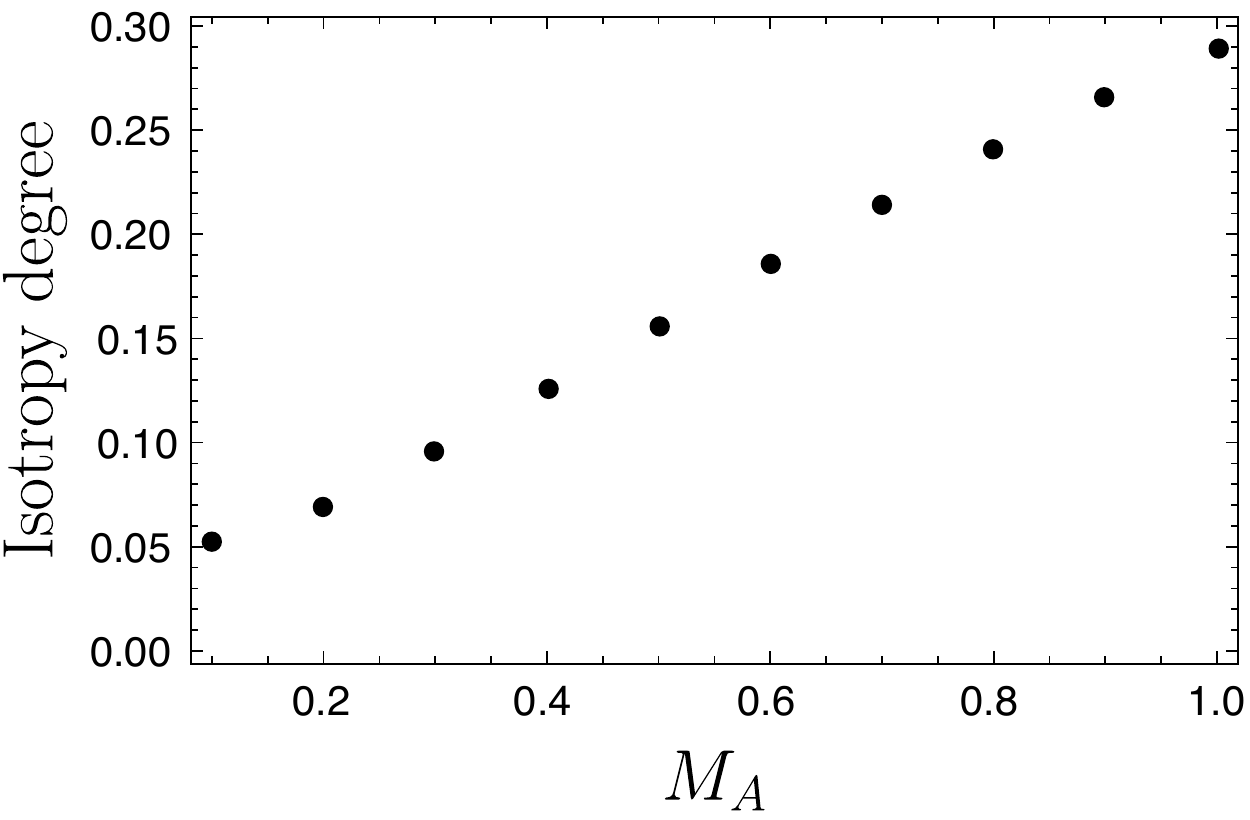}
\caption{Alfv\'en mode. Left-hand and center: quadrupole to monopole and octupole to monopole ratio for various $\gamma$. Solid line is for $M_{\text{A}}=0.1$, dotted line for $M_{\text{A}}=0.4$ and dashed line for $M_{\text{A}}=0.7$. Right-hand: isotropy degree for various $M_{\text{A}}$ at $\gamma=\upi/2$.}
\label{fig4}
\end{figure*}

Fig. \ref{fig4} shows some important properties of Alfv\'en mode. Firstly, looking at this figure one can clearly see that this mode becomes more isotropic with increasing Alfv\'en Mach number $M_{\text{A}}$, as characterised by the decreasing level of quadrupole to monopole and octupole to monopole ratio and increasing level of isotropy degree with increasing $M_{\text{A}}$. It is also quite clear from the left-hand and central panel of the figure that this mode becomes highly anisotropic at $\gamma=\upi/2$, which is expected. Note that the finite quadrupole to monopole ratio at $\gamma=0$ is misleading in a sense that both quadrupole and monopole vanish at $\gamma=0$. In the case when one considers the mixture of modes, this problem is remedied as slow modes and high-$\beta$ fast modes have non-vanishing monopole at $\gamma=0$.

\subsection{Slow mode}
Slow modes in high-$\beta$ plasma are similar to pseudo-Alfv\'en modes in incompressible regime, while at low-$\beta$ they are density perturbations propagating with sonic speed parallel to magnetic field (see \citealt{cho2003compressible}). The power spectrum of this mode is the same as that of Alfv\'en mode.
Slow modes in high-$\beta$ is purely $F$-type and therefore, the correlation is given by (see \citetalias{kandel2016extending})
\begin{equation}
\left(\hat{\xi}_{\bm{k}} \otimes \hat{\xi}^*_{\bm{k}}\right)_{ij}=\frac{(\hat{\bm{k}}\cdot\hat{\lambda})^2\hat{k}_i\hat{k}_j+\hat{\lambda}_i\hat{\lambda}_j-(\hat{k}\cdot\hat{\lambda})(\hat{k}_i\hat{\lambda}_j+\hat{k}_j\hat{\lambda}_i)}{1-(\hat{\bm{k}}\cdot\hat{\lambda})^2}~.
\end{equation}
With this, we can write
\begin{equation}\label{fast2dcorr}
\left(\hat{\xi}_{\bm{K}} \otimes \hat{\xi}^*_{\bm{K}}\right)_{zz}=\frac{\hat{\lambda}_z\hat{\lambda}_z}{1-(\hat{\bm{K}}\cdot\hat{\Lambda})^2}=\frac{\cos^2\gamma}{1-\sin^2\gamma\cos^2\psi}~.
\end{equation}
Using Eqs. \eqref{centriodcoeff} and \eqref{fast2dcorr}, one can obtain the multipole moments of the centroid structure function as
\begin{equation}\label{slowstrdn}
\mathcal{D}_n(R)=C_n(2/3)\sum_{p=-\infty}^{\infty}\hat{A}_p\mathcal{W}_{n-p}^SR^{5/3}~,
\end{equation}
where the spectral weight function $\mathcal{W}_{p}^S$ is given by
\begin{align}\label{slowspect}
\mathcal{W}_{n-p}^S=\frac{1}{2\upi}\int_{0}^{2\upi}\mathop{\mathrm{d}\psi}\mathrm{e}^{-\mathrm{i}(n-p)\psi}\frac{\cos^2\gamma}{1-\cos^2\psi\sin^2\gamma}\nonumber\\
=\cos\gamma\left(\frac{1-\cos\gamma}{\sin\gamma}\right)^{|n-p|}~.
\end{align}
The spectral weight function of slow modes, given by Eq. \eqref{slowspect}, is plotted in the left-hand panel of Fig. \ref{fig3}. It is clear from Fig. \ref{fig3} that $\mathcal{W}_{n-p}^S$ vanishes at $\gamma=\upi/2$ for all $n-p$, and therefore, the structure function vanishes at this angle. In the opposite case, $\gamma=0$, $\mathcal{W}_{n-p}^S$ vanishes for all non-zero $n-p$ and equal to 1 for $n=p$, but $\hat{\mathcal{A}}_p=0$ for $p>0$; therefore, no anisotropy is present. For general $\gamma$, $\mathcal{W}_{n-p}^S$ decays very rapidly with increasing $|n-p|$, which implies that for  practical purposes, it is enough to just few terms near $p\approx n$ in the sum presented in Eq. \eqref{slowstrdn}.

Slow modes in low-$\beta$ have their correlation function as $\langle v_iv_j\rangle\propto \hat{\lambda}_i\hat{\lambda}_j$ and therefore, it can be straightforwardly shown that 
\begin{equation}\label{slowlowdn}
\mathcal{D}_n(R)=C_n(2/3)\cos^2\gamma \mathcal{\hat{A}}_n R^{5/3}~.
\end{equation}

\begin{figure*}
\centering
\includegraphics[scale=0.4]{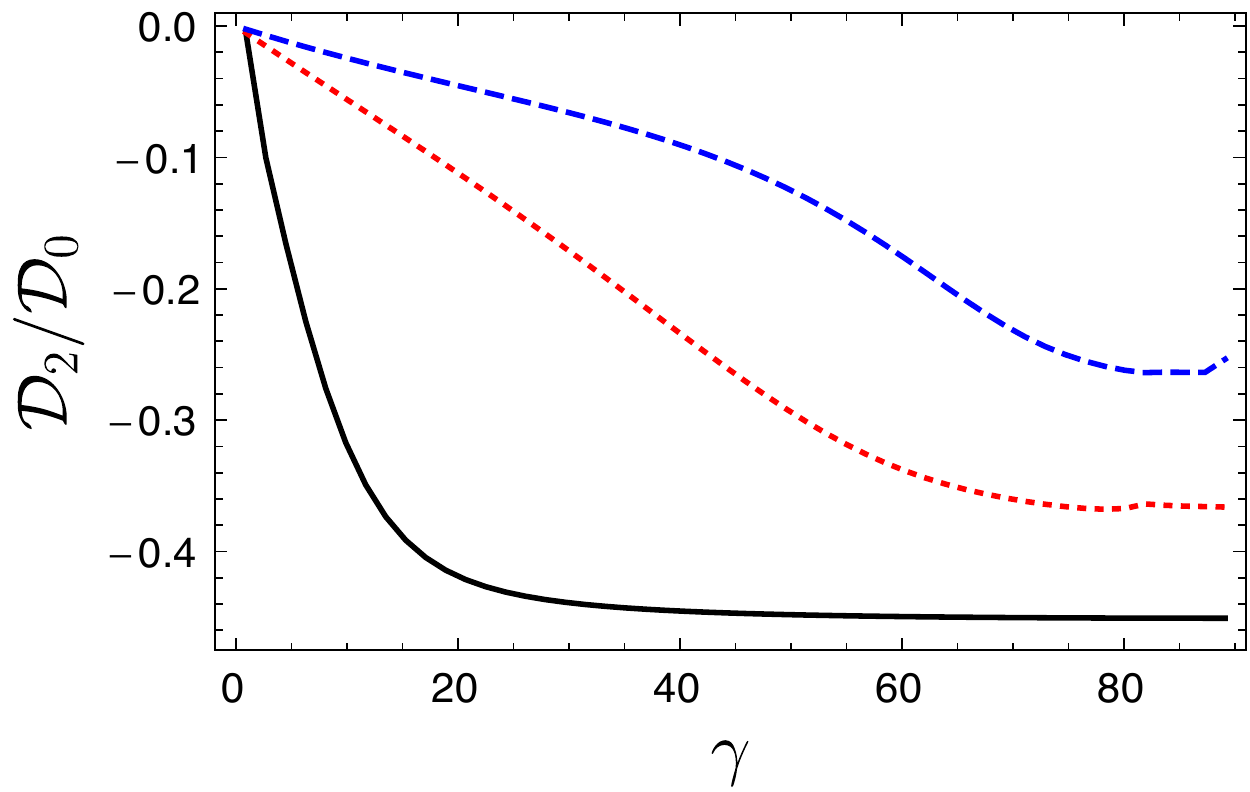}\hspace*{0.2cm}
\includegraphics[scale=0.4]{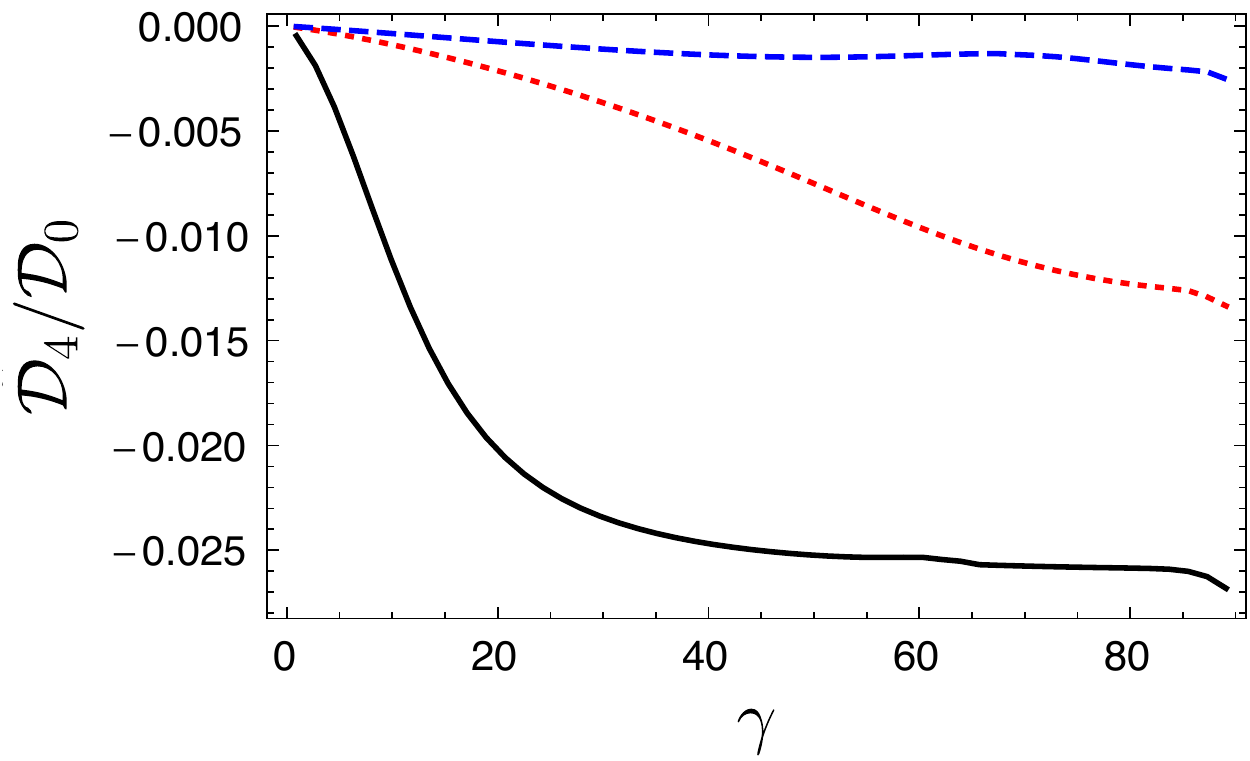}\hspace*{0.2cm}
\includegraphics[scale=0.4]{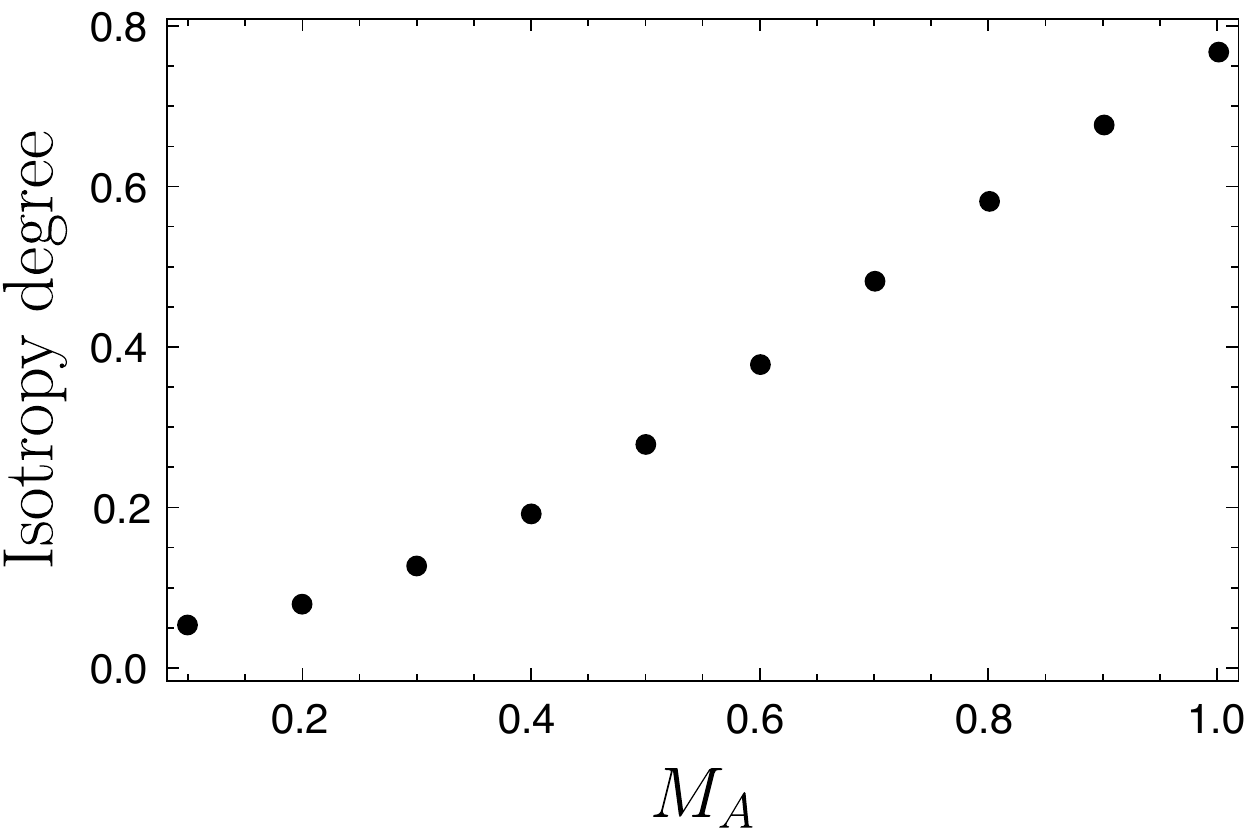}
\caption{High-$\beta$ slow mode. Left-hand and center: quadrupole to monopole and octupole to monopole ratio for various $\gamma$. Solid line is for $M_{\text{A}}=0.1$, dotted line for $M_{\text{A}}=0.4$ and dashed line for $M_{\text{A}}=0.7$. Right-hand: isotropy degree for various $M_{\text{A}}$ at $\gamma=\upi/3$.}
\label{fig6}
\end{figure*} 

\begin{figure*}
\centering
\includegraphics[scale=0.4]{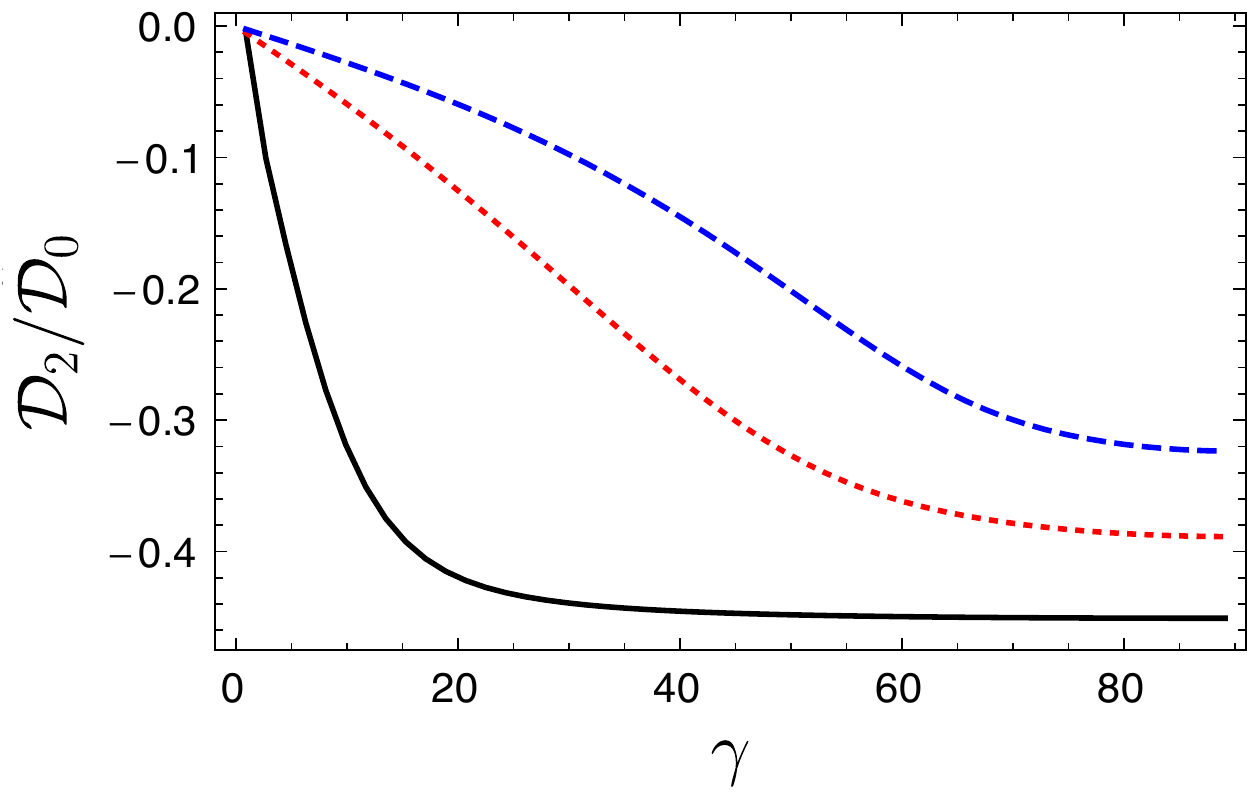}\hspace*{0.2cm}
\includegraphics[scale=0.4]{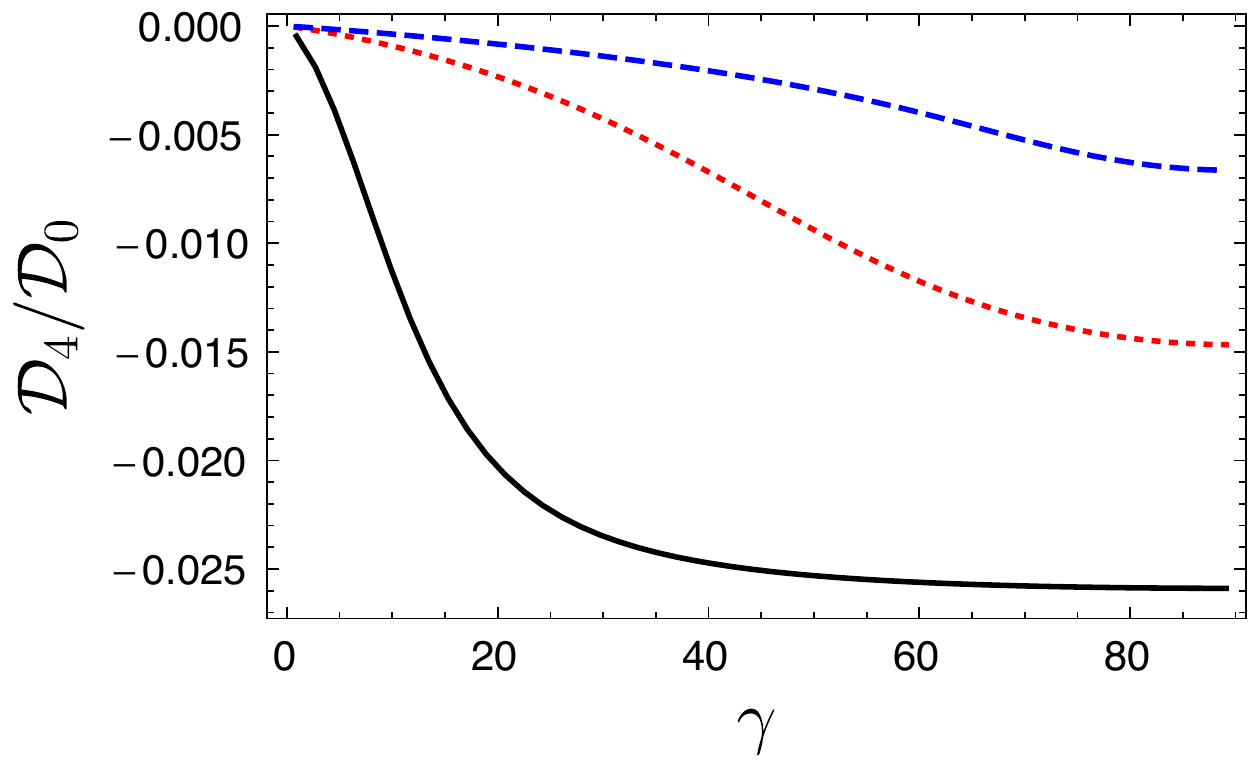}\hspace*{0.2cm}
\includegraphics[scale=0.4]{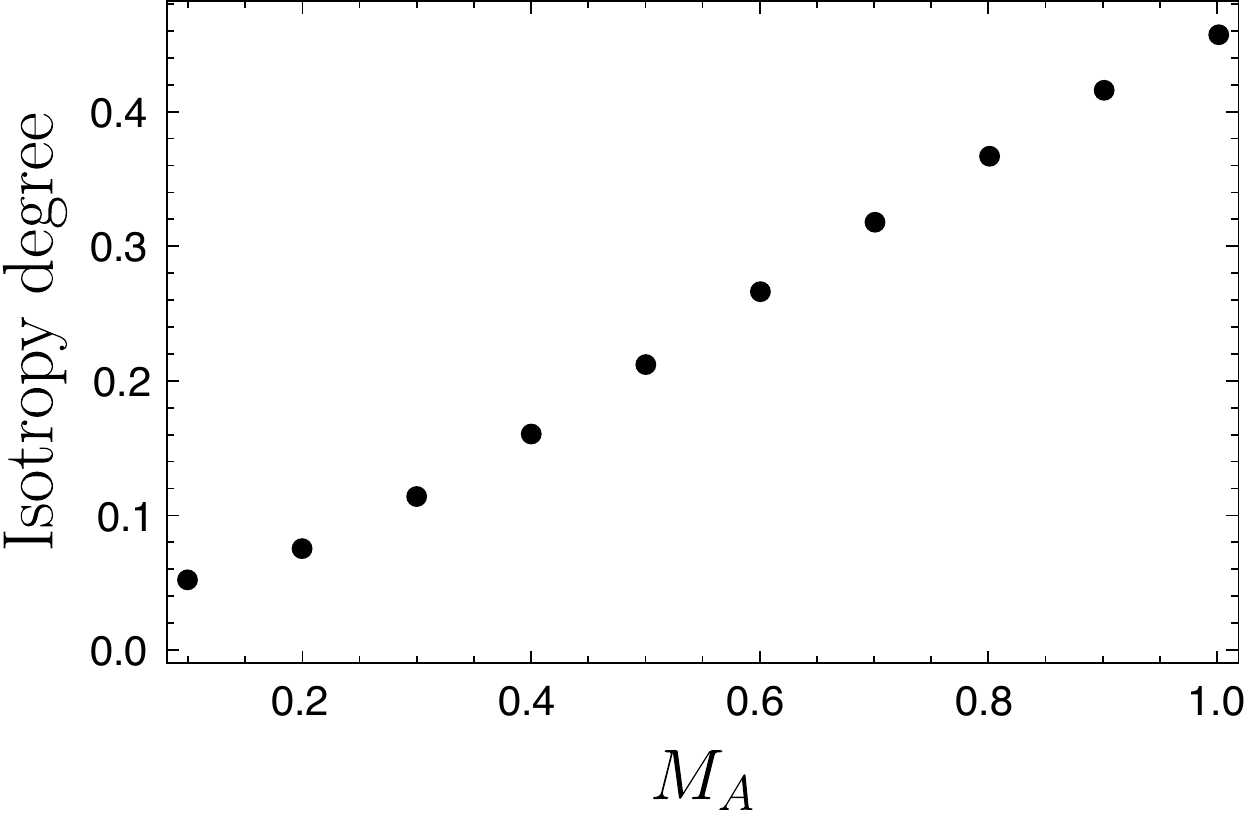}
\caption{Low-$\beta$ slow mode. Left-hand and center: quadrupole to monopole and octupole to monopole ratio for various $\gamma$. Solid line is for $M_{\text{A}}=0.1$, dotted line for $M_{\text{A}}=0.4$ and dashed line for $M_{\text{A}}=0.7$. Right-hand: isotropy degree for various $M_{\text{A}}$ at $\gamma=\upi/3$.}
\label{fig7}
\end{figure*} 

Slow modes in both high- and low-$\beta$ plasma are  highly anisotropic at small $M_{\text{A}}$ and become more isotropic with increasing $M_{\text{A}}$. This is clearly shown in Figs. \ref{fig6} and \ref{fig7}. Moreover, the anisotropy level of both high- and low-$\beta$ slow modes are similar. This is because the dominant term in Eq.\eqref{slowstrdn} is the diagonal term $n=p$, while the $\cos^2\gamma$ term in Eq. \eqref{slowlowdn} cancels upon taking ratio of multipole moments, thus the ratio of multipole moments in both cases yield similar results. It is important to note that the anisotropy of slow modes at $\gamma=\upi/2$ cannot be measured as both quadrupole and monopole vanish at $\gamma=\upi/2$.

\subsection{Fast mode}
Fast modes in high-$\beta$ are purely compressible modes with a velocity tensor structure given by 
\begin{equation}
\left(\hat{\xi}_{\bm{k}} \otimes \hat{\xi}^*_{\bm{k}}\right)_{ij}=\hat{k}_i\hat{k}_j~.
\end{equation}
The power spectrum of fast modes is isotropic, and therefore the velocity correlation tensor is isotropic as well. 

\begin{figure*}
\centering
\includegraphics[scale=0.4]{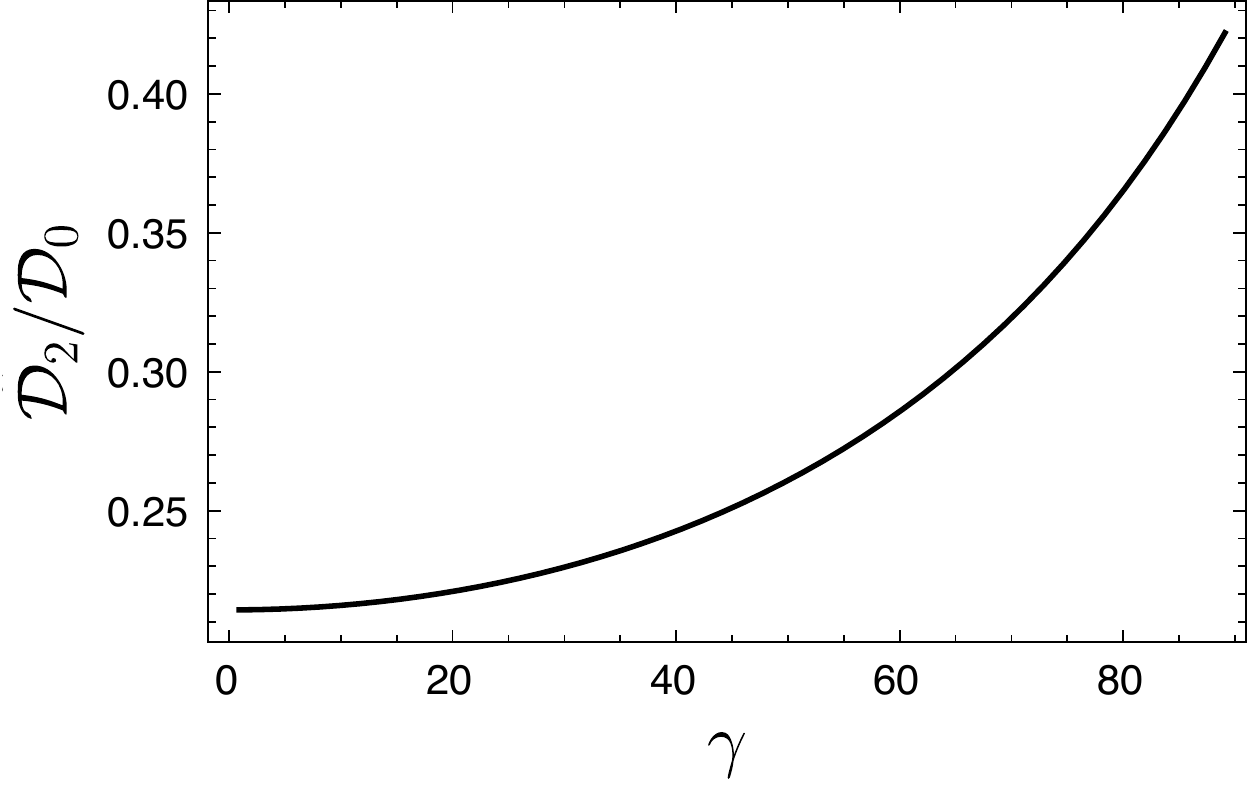}\hspace*{0.2cm}
\includegraphics[scale=0.4]{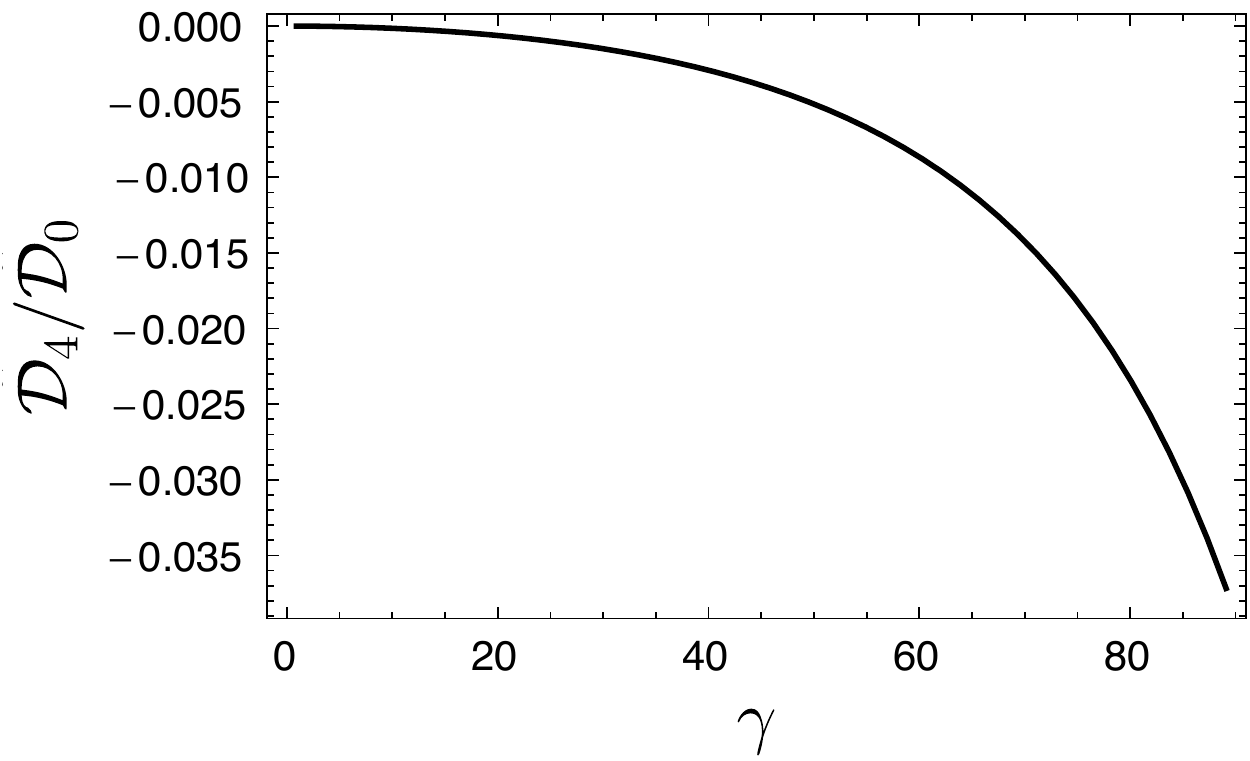}\hspace*{0.2cm}
\includegraphics[scale=0.4]{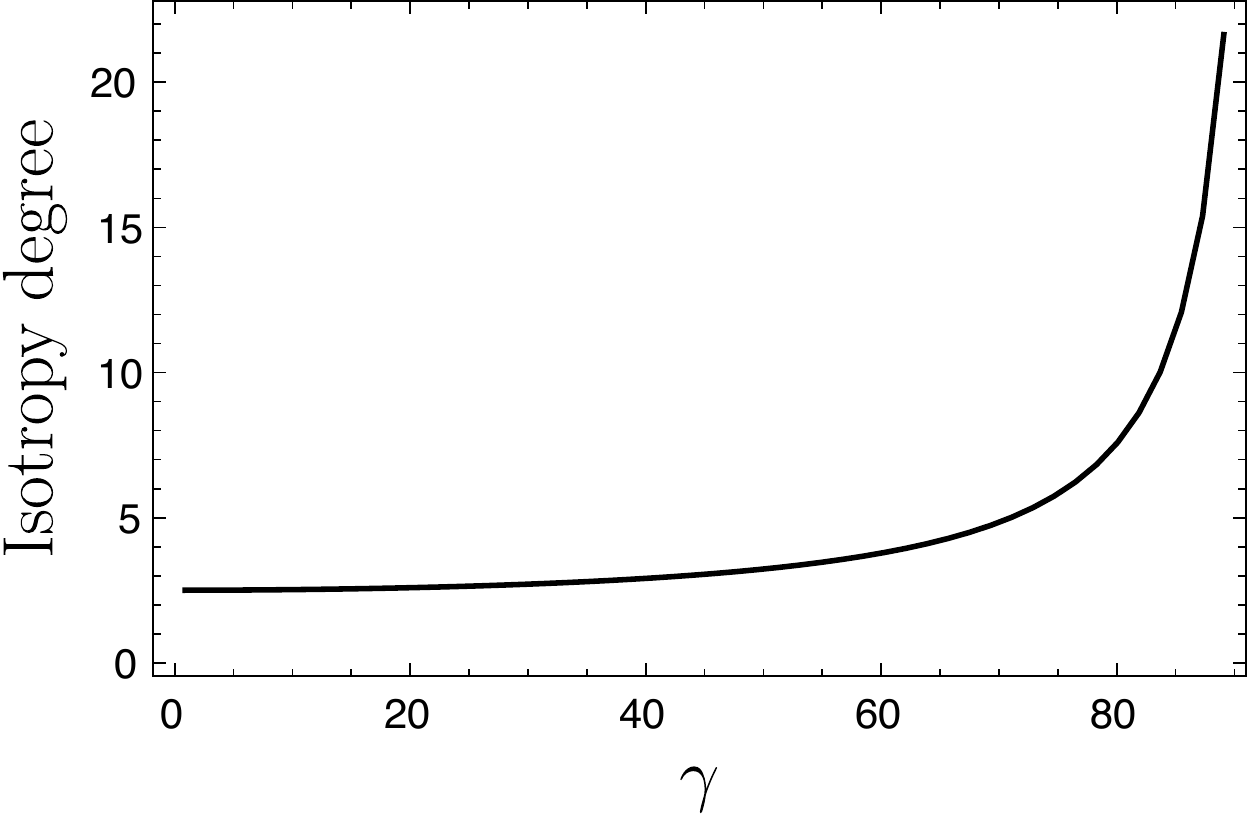}
\caption{Low-$\beta$ fast mode. Left-hand to right-hand: quadrupole to monopole, octupole to monopole ratio and isotropy degree for various $\gamma$.}
\label{fig5}
\end{figure*}

On the other hand, low-$\beta$ fast mode is anisotropic with the anisotropy built in the tensor (see \citetalias{kandel2016extending})
\begin{equation}\label{fastcormode}
\left(\hat{\xi}_{\bm{k}} \otimes \hat{\xi}^*_{\bm{k}}\right)_{ij}=\frac{\hat{k}_i\hat{k_j}-(\hat{k}.\hat{\lambda})(\hat{k}_i\hat{\lambda}_j+\hat{k}_j\hat{\lambda}_i)+(\hat{k}.\hat{\lambda})^2\hat{\lambda}_i\hat{\lambda}_j}{1-(\hat{k}.\hat{\lambda})^2}~.
\end{equation}
Making use of Eq. \eqref{fastcormode}, we obtain
\begin{equation}
\left(\hat{\xi}_{\bm{K}} \otimes \hat{\xi}^*_{\bm{K}}\right)_{zz}=\frac{(\sin\gamma\cos\gamma)^2\cos^2\psi}{1-\sin^2\gamma\cos^2\psi}~.
\end{equation}
Keeping in mind that fast modes have isotropic power spectrum so that only $\hat{\mathcal{A}}_0$ in non-vanishing, we have
\begin{equation}\label{fastnthcorr}
\mathcal{D}_n(R)=C_n(1/2)\mathcal{A}_0\mathcal{W}^F_{n}R^{3/2}~,
\end{equation} 
where the spectral weight function $W_n^F$ is defined as
\begin{align}
\mathcal{W}_{n}^F=\frac{1}{2\upi}\int_0^{2\upi}\mathop{\mathrm{d}\psi}&\mathrm{e}^{-\mathrm{i}n\psi}\frac{(\sin\gamma\cos\gamma)^2\cos^2\psi}{1-\sin^2\gamma\cos^2\psi}\nonumber\\
&=-\cos^2\gamma\delta_{n0}+\cos\gamma\left(\frac{1-\cos\gamma}{\sin\gamma}\right)^{|n|}~.
\end{align}
This spectral weight function of fast mode is plotted in the central panel of Fig. \ref{fig3}, which shows that this function vanishes both at $\gamma=0$ and  $\gamma=\upi/2$. The left-hand panel of Fig. \ref{fig5} shows that the quadrupole to monopole ratio of low-$\beta$ fast mode is $\sim 0.3$ through out the entire range of $\gamma$. The quadrupole to monopole ratio somewhat increases with increasing $\gamma$  to its maximum value $\approx 0.4$ at $\gamma=\upi/2$, however, the amplitude of both monopole and quadrupole is $\sim 0$ at $\gamma\sim \upi/2$. In fact, the optimal signal is obtained at $\gamma\sim \upi/3$.  

Note that since $C_2(1/2)>0$, the quadrupole moment of fast mode is positive, which is also distinct from Alfv\'en mode. We found that this is due to the fact that anisotropy of fast mode comes from its anisotropic tensor structure and not from its power spectrum. 

\subsection{Mixture of modes}
\begin{figure*}
\centering
\includegraphics[scale=0.4]{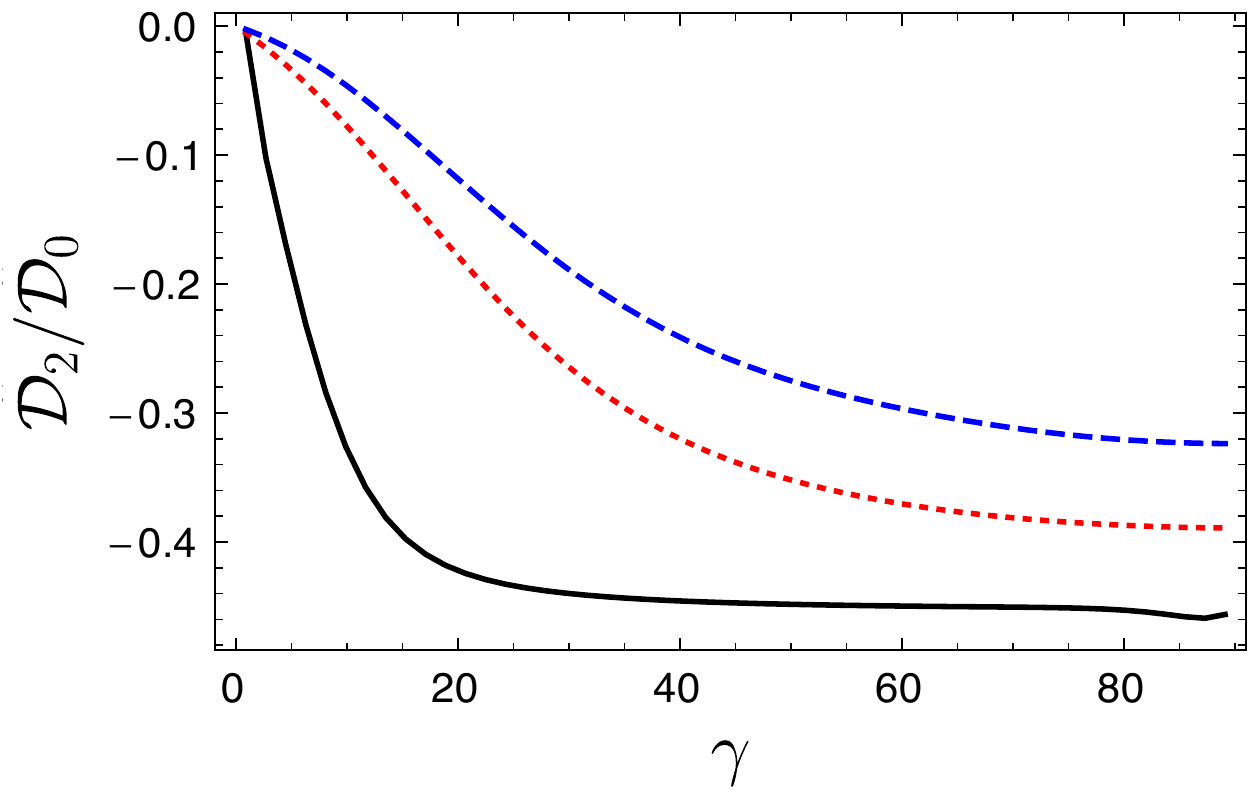}\hspace*{0.2cm}
\includegraphics[scale=0.4]{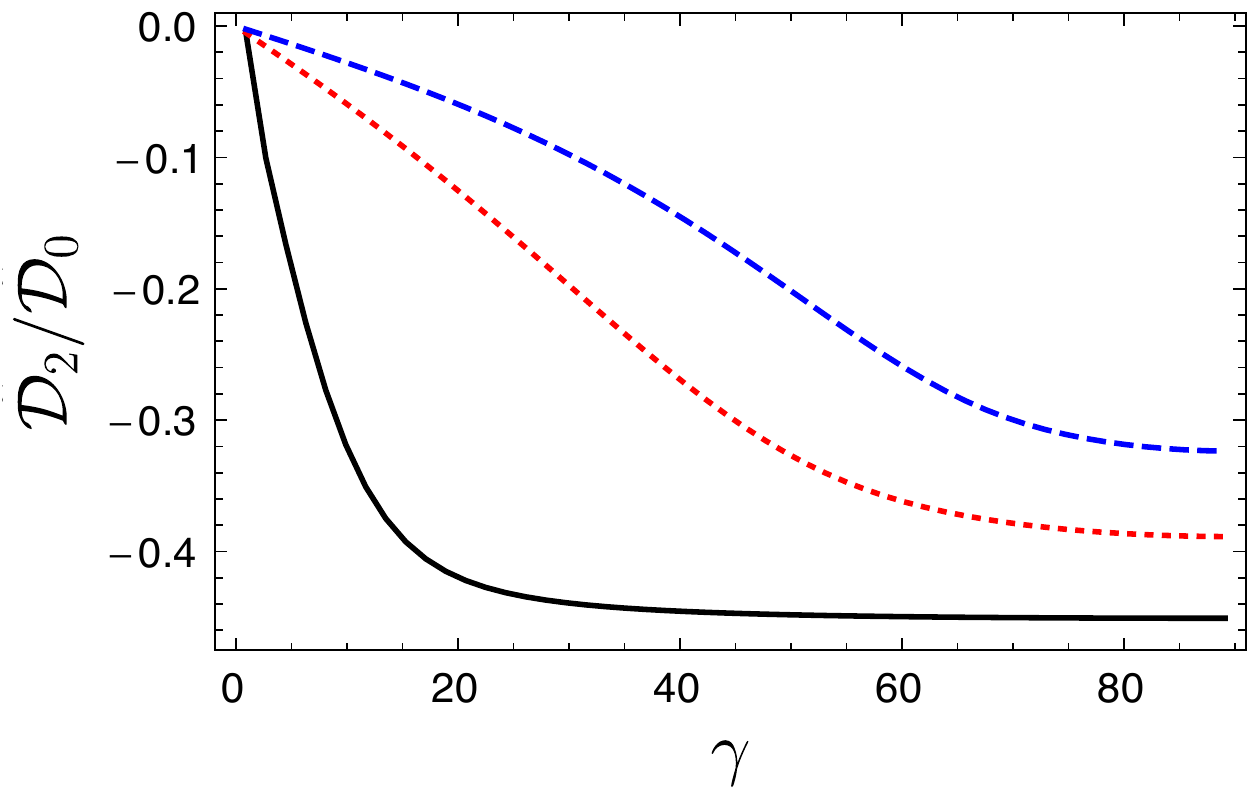}\hspace*{0.2cm}
\caption{Mixture of modes. Left-hand: quadrupole to monopole ratio for a mixture of 85\% Alfv\'en and 15\% high-$\beta$ slow modes. Right-hand: same for a mixture of 50\% Alfv\'en and 50\% low-$\beta$ slow modes. Solid line is for $M_{\text{A}}=0.1$, dotted line for $M_{\text{A}}=0.4$ and dashed line for $M_{\text{A}}=0.7$.}
\label{fig8}
\end{figure*}

Real world setting of MHD turbulence involves superposition of the different MHD modes. Therefore, we consider the effect of  mixtures of different MHD modes in the observed centroids anisotropy. In the case of mixture between Alfv\'en and slow  modes, Fig. \ref{fig8} clearly shows that the observed anisotropy is unaffected by this mixture in different regimes. For  instance at $\gamma\gtrsim\upi/4$ the observed anisotropy of the mixture is the same as that of Alfv\'en mode alone while at $\gamma\lesssim\upi/4$, the anisotropy level is similar to that of slow modes alone. This is again due to the fact that at $\gamma\approx \upi/2$ signal from Alfv\'en mode is dominant while at $\gamma\approx 0$ signal from slow mode is dominant. On the other hand, we  expect the mixture of fast mode with other two modes to decrease the level of anisotropy. This is because the quadrupole moment (which is the measure of anisotropy) of fast mode is opposite in sign than that of other modes.

\subsection{Density effects}
\begin{figure*}
\includegraphics[scale=0.4]{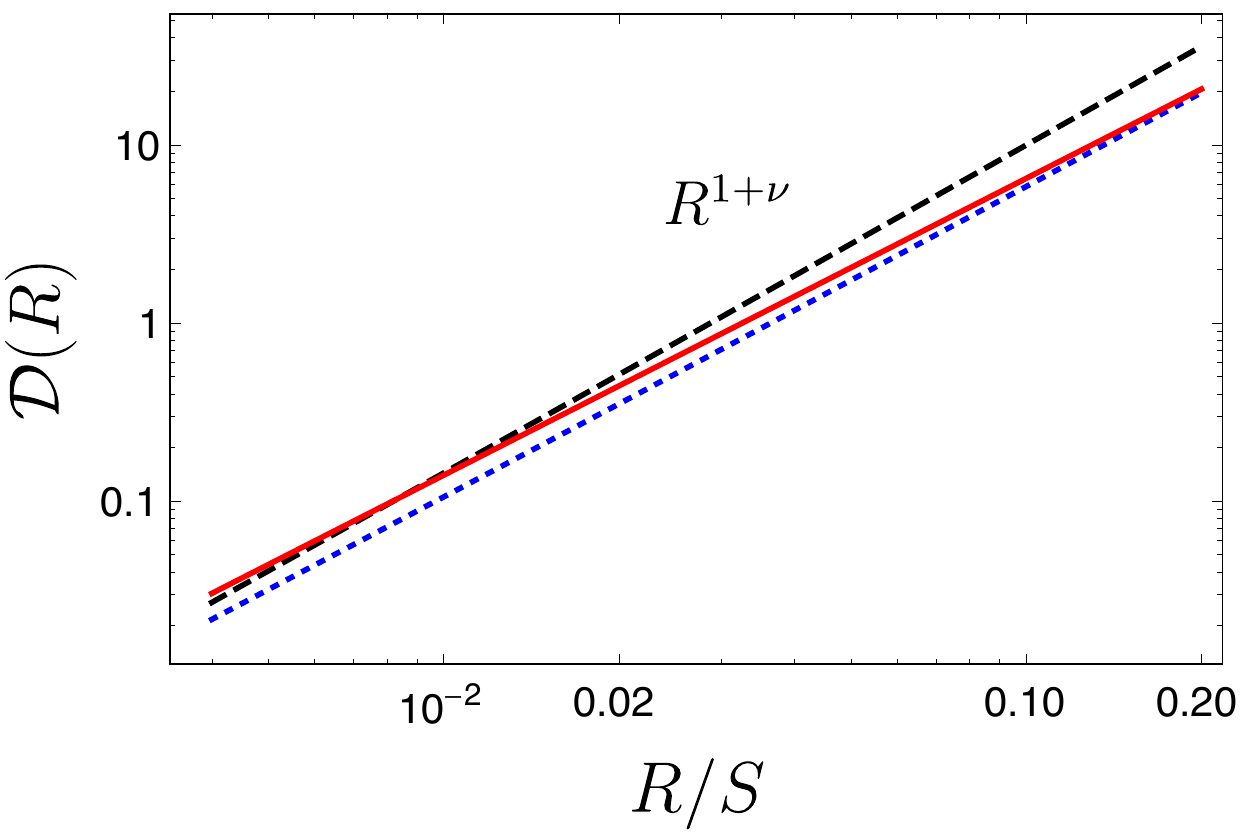}\hspace*{0.2cm}
\includegraphics[scale=0.4]{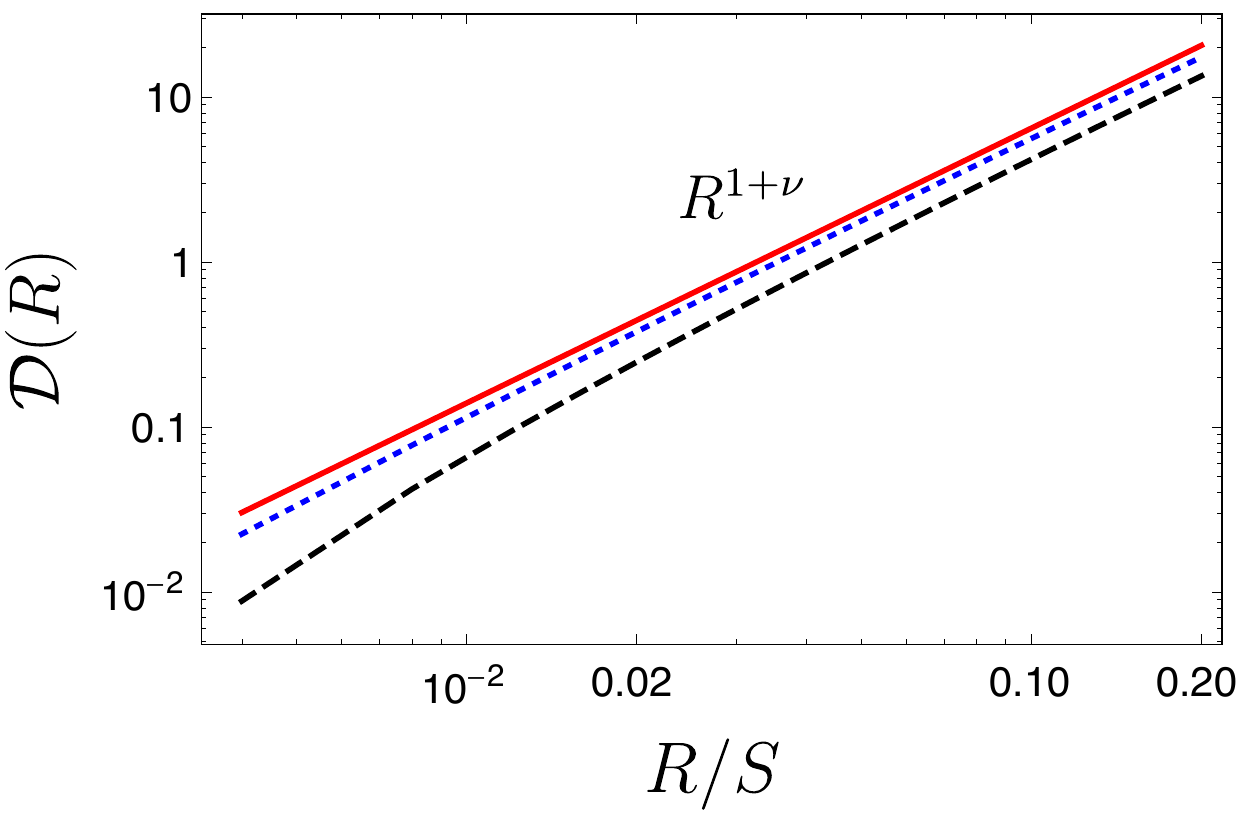}\hspace*{0.2cm}
\caption{First panel: MVC structure function for various $\sigma_\rho/\rho_0$ for {\it steep} velocity field of Kolmogorov index $2/3$ and density fields of index $\nu_\rho=1/2$. The solid line is expected power law of $R^{5/3}$, the dotted line is for $\sigma_\rho/\rho_0=0.5$ and the dashed line for $\sigma_\rho/\rho_0=1$. One can see some deviation from the power law behaviour already at $\sigma_\rho/\rho_0=1$. This deviation is expected to be stronger with increasing $\sigma_\rho/\rho_0$. Second panel: The same but for {\it shallow} density with $\nu_\rho=-1/2$. The dotted line is for $\sigma_\rho/\rho_0=0.5$ and dashed line is for $\sigma_\rho/\rho_0=1$. Solid line is the power law $R^{5/3}$ from pure velocity effects.}
\label{fig9}
\end{figure*}

The main aim of using velocity centroids is to obtain information about velocity spectrum. Looking at Eq. \eqref{correlation}, one can see that the centroid structure function contains not only the contribution from velocity effects but also from density effects. In this regard, separating velocity contribution from density contribution is not always possible. In particular, if the density spectrum is shallow, as is the case for super-sonic turbulence, one might not be able obtain the velocity spectra from the centroids.  On the other hand, for a steep density spectra the velocity spectra can be extracted if the density dispersion  in a turbulent field is less than the mean density \citep{esquivel2007statistics}. This has been clearly illustrated in the left-hand panel of the Fig. \ref{fig9}, where the centroid structure function is plotted for various ratios of $\sigma_\rho/\rho_0$. It is clearly shown in the figure that the velocity spectra can be obtained when $\sigma_\rho/\rho_0<1$, while the spectra is corrupted by the density-velocity cross-term when $\sigma_\rho/\rho_0>1$. 

As explained in \citet[henceforth \citetalias{lazarian2003statistics}]{lazarian2003statistics}, centroids can trace the velocity spectrum if the centroid structure function is much larger than the first term of Eq. \eqref{correlation}. If this condition is not fulfilled, the velocity spectrum can be obtained by subtracting the first term (product of velocity dispersion and density correlation) of Eq. \eqref{correlation} as long as the density correlation can be measured independently\footnote{Note that one can obtain this contribution from the density term observationally by measuring intensity fluctuations. For instance, in the language of VCA this term can be obtained through intensity statistics in the `thick slice' limit (see \citetalias{lazarian2000velocity} for more details).} and density-velocity correlation is not strong. A potential challenge is the determination of velocity dispersion in the case when thermal broadening is large. However, one can circumvent this by using emission lines of heavier species. In this regard, \citetalias{lazarian2003statistics} introduced a notion of `modified' velocity centroids (MVC), where the first term of Eq. \eqref{correlation} was subtracted. Formally, the structure function of MVC in the absence of density-velocity correlation is 
\begin{equation}\label{modifiedcorrelation}
\mathcal{D}_{\text{MVC}}(\bm{R})\approx\int_{-S}^S\mathop{\mathrm{d}z}[\xi_\rho(\bm{r})D_z(\bm{r})-\xi_\rho(0,z)D_z(0,z)]~.
\end{equation}

\begin{figure*}
\includegraphics[scale=0.4]{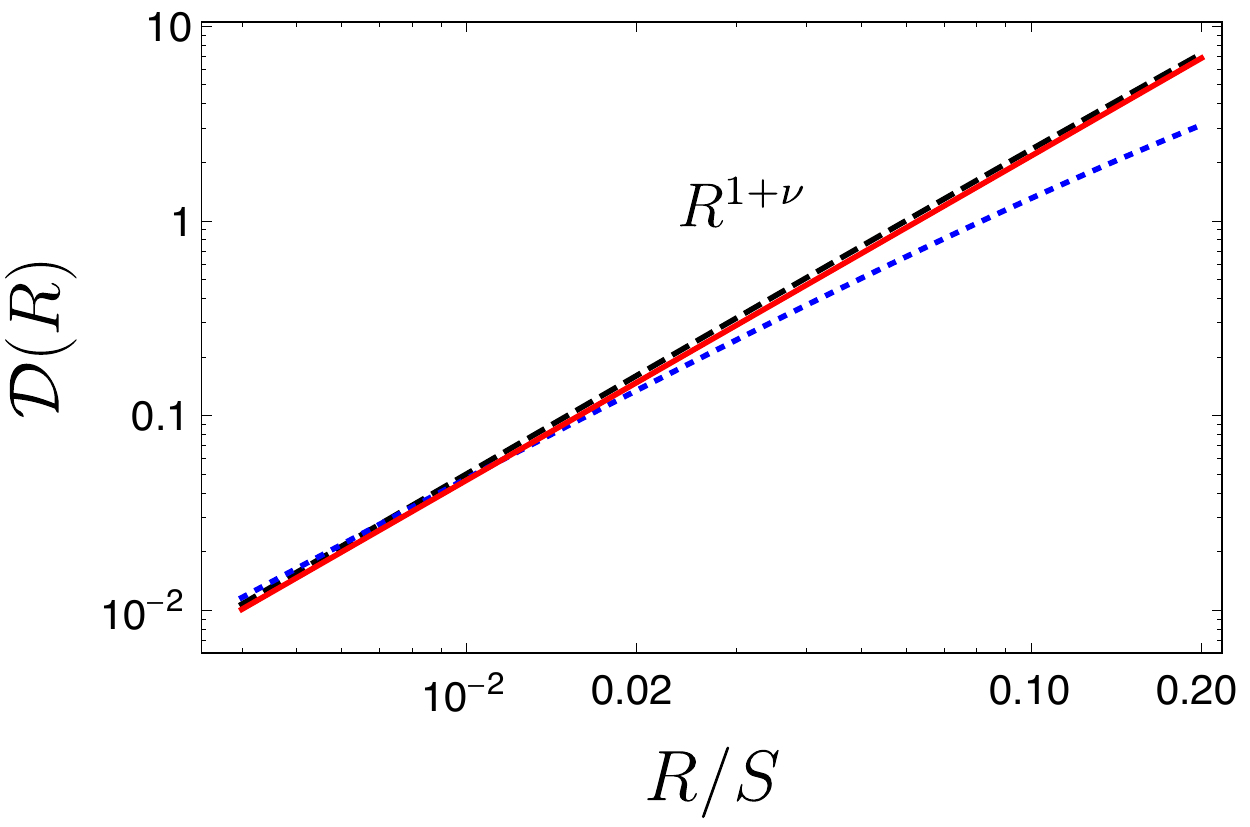}\hspace*{0.2cm}
\includegraphics[scale=0.4]{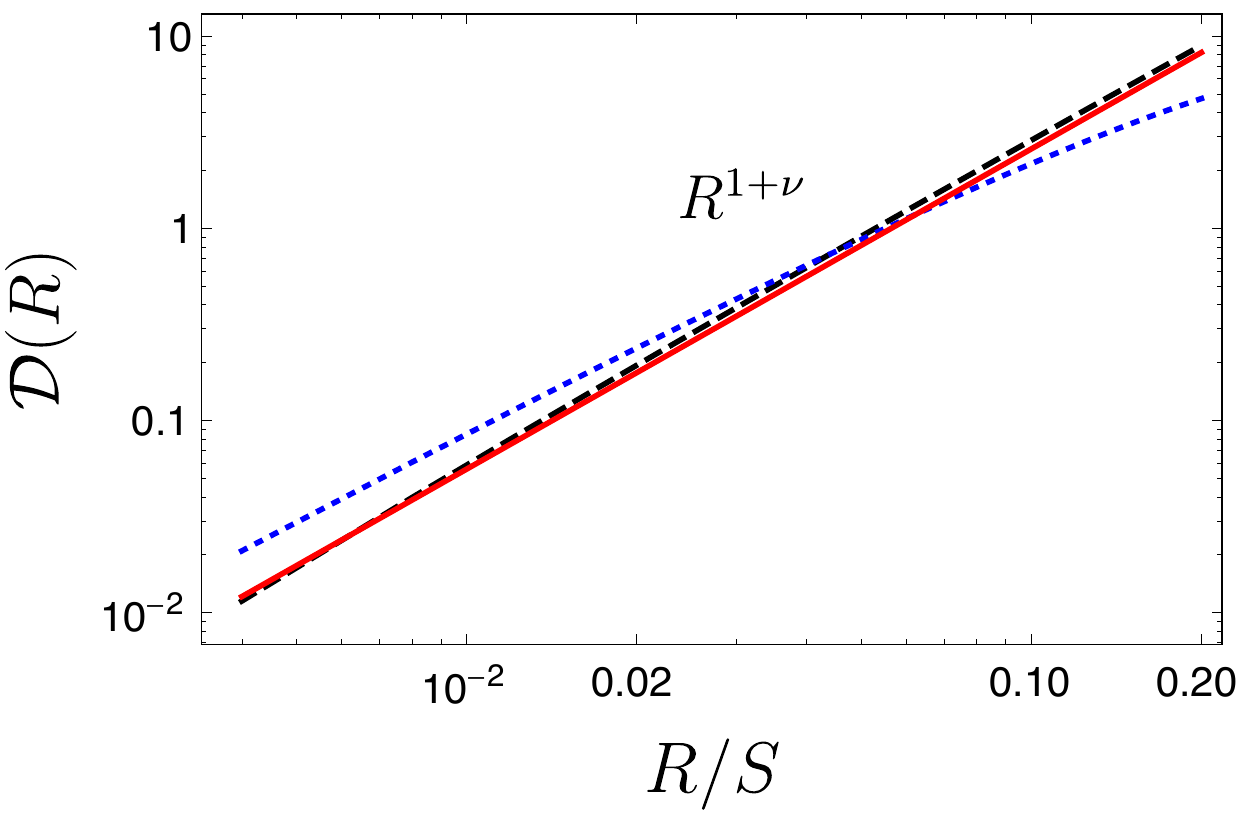}
\caption{Plot showing comparison of MVC with UVC at $\sigma_\rho/\rho_0=0.5$ at short scales $R<S$. The dashed line in both the panels are for MVC, dotted line is for UVC and solid line shows the power law $R^{5/3}$ from pure velocity effects. In both figures a {\it steep} velocity spectrum of Kolmogorov index is assumed, whereas density spectrum is assumed to be {\it steep}  (with $\nu_\rho=1/2$) in the first panel and {\it shallow} (with $\nu_\rho=-1/2$) in the second. It is clear that MVC works well for both steep and shallow spectra.}
\label{fig10}
\end{figure*}

It was explained in LE06 that MVCs can trace the velocity spectrum better than the `unnormalised' centroids (UVC) if the lag $R$ under study be smaller than the saturation scale of the velocity structure function as well as the line of sight extent $S$ of the turbulent cloud. We find that the MVC is able to trace the velocity spectra even for a shallow spectrum, as illustrated in Fig. \ref{fig9}. In fact, Fig. \ref{fig10} clearly show that modified centroids work better than UVC at smaller lags $R$. Note that for shallow density field, density-velocity cross-term yields a scaling $R^{1+\nu_\rho+\nu}$, while pure velocity term yields $R^{1+\nu}$, and since $\nu_\rho<0$ for a shallow density spectra, this cross-term scaling can dominate the MVC scaling extremely small scale. Although, we see that MVCs work well even for shallow density spectra, there are two important points to make. First, we require $\sigma_\rho/\rho_0<1$ to obtain velocity spectra correctly, otherwise density-velocity correlation (which we ignored) becomes important (\citealt{esquivel2007statistics}). However, shallow density often does not fulfil this criteria. Second, shallow density is often associated with high sonic Mach number $M_{\text{s}}$ where non-Gaussian features are often prominent, significantly affecting the statistics. To sum up, one needs to know if $\sigma_\rho/\rho_0$ to conclude if MVCs work for shallow spectra.

Our study of anisotropy was for constant density field. In the case when density field is anisotropic, one should also account for the anisotropy due to density effects as well. For MVC the anisotropy is dominated by velocity effects as long as the density dispersion is less than the mean density. On the other hand, both density and velocity effects contribute to the UVC anisotropy.

\subsection{Comparisons with earlier numerical works}
The numerical study of anisotropies with centroids has been carried out in the past in LE06, \citet[hereafter \citetalias{esquivel2011velocity}]{esquivel2011velocity} and \citet[hereafter \citetalias{burkhart2014measuring}]{burkhart2014measuring}. Here we compare our findings with the findings of \citetalias{esquivel2011velocity} and \citetalias{burkhart2014measuring}. \citetalias{esquivel2011velocity} studied anisotropies at $\gamma=\upi/2$, while \citetalias{burkhart2014measuring} studied anisotropy at varying $\gamma$ as well. Both of these studies found out a clear dependence of anisotropy with Alfv\'en Mach number $M_{\text{A}}$. They reported that the anisotropy increases with decreasing $M_{\text{A}}$, which coincides with our result. As an example, at $M_{\text{A}}=0.7$ and $M_{\text{s}}=2.3$, the degree of isotropy in both the papers was reported to be $\sim 0.3$, while our results show that in the case when Alfv\'en and slow modes are dominant, the isotropy degree at $M_{\text{A}}=0.7$ is around $\sim 0.25$, which is close to their results.  Our finding that the degree of anisotropy is highest at $\gamma=\upi/2$ matches  with the findings in \citetalias{burkhart2014measuring}, where it was stated that at $\gamma=\upi/2$ the anisotropy is highest regardless of the sonic Mach number $M_{\text{s}}$. It is clear from our finding that the isotropy degree of centroid is clearly dependent on $M_{\text{A}}$. However, it was discussed in \citetalias{esquivel2011velocity} and \citetalias{burkhart2014measuring} that besides a dependence of isotropy degree of $M_{\text{A}}$, there exists a weak dependence on $M_{\text{s}}$ as well. Although there is no direct role of $M_{\text{s}}$ in determining the degree of isotropy in our formalism, this weak dependence can be explained by noting that with an increasing $M_{\text{s}}$, there is an increasing contribution from fast mode, thus decreasing the level of anisotropy. This is consistent with the results in \citetalias{esquivel2011velocity} and \citetalias{burkhart2014measuring}.

\subsection{Effects of self-absorption in anisotropy studies}
In Sections \ref{sec:genanisotropy} and \ref{sec:mhdmodes}, we studied anisotropy for an optically thin medium. With the theory developed in Sec. \ref{othick}, it is straightforward to extend the study of anisotropy to optically thick media. 

The extent of self-absorption sets an effective  cut-off in velocity difference beyond which signals does not contribute to the correlation, and thus we may expect changes in anisotropy level as this extent changes. Our study in Sec. \ref{othick} suggests that in the case of weak self-absorption, the centroid structure function behaves similar to that of optically thin regime, while at stronger self-absorption it behaves as thin-slice regime of VCA (see \citetalias{lazarian2000velocity}). Naturally,  in this case of weak self-absorption one should see the level of anisotropy similar to the optically thin case of usual centroids, while at stronger self-absorption the anisotropy level will decrease to the level of thin-slice regime of VCA.  For example, the isotropy degree of Alfv\'en mode at $M_A=0.7$ and at $\gamma=\upi/2$ changes from 0.25 at weak absorption to 0.68 at strong absorption. 

\begin{figure}
\includegraphics[scale=0.5]{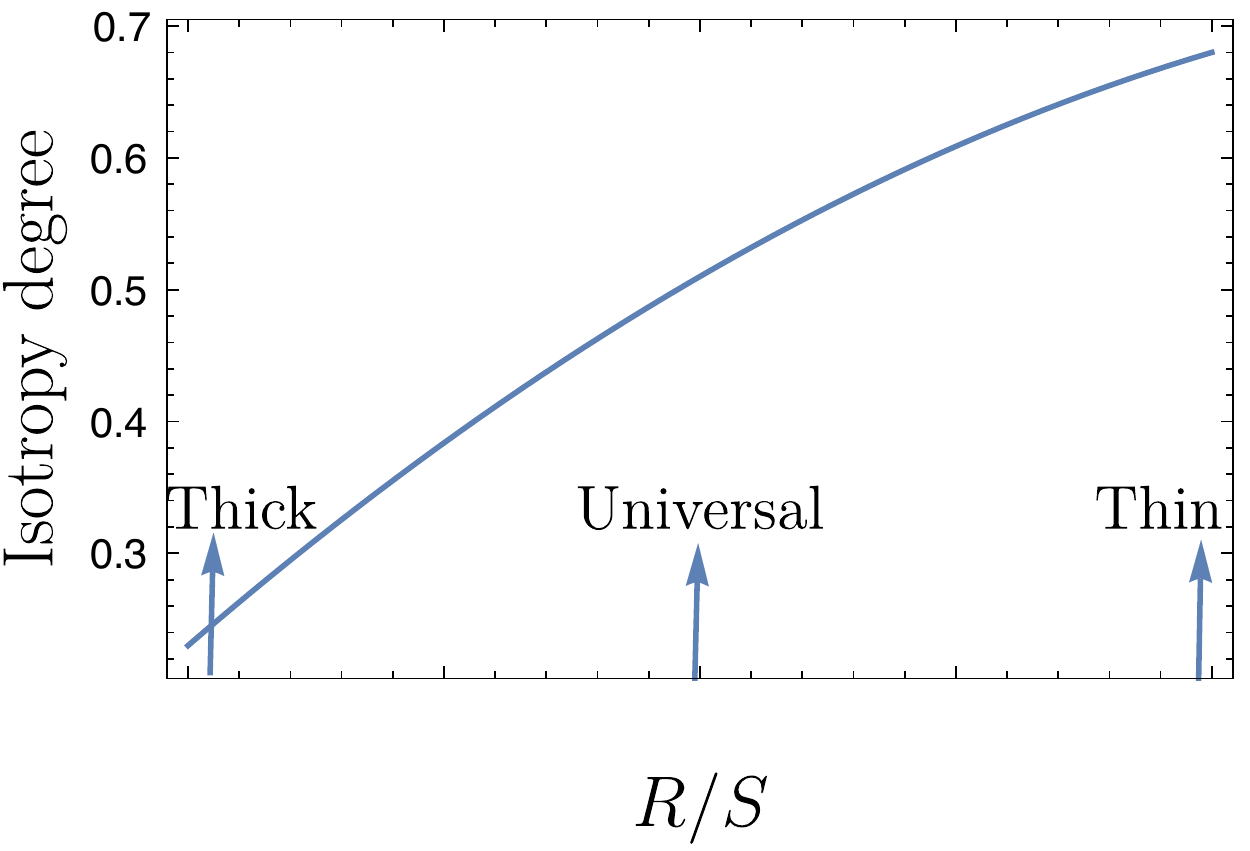}
\caption{Expected change of isotropy degree at various scales $R$ in the presence of self-absorption for Alfv\'en mode at $M_A=0.7$ and at $\gamma=\upi/2$. The curve is produced by using three theoretically predicted values of isotropy degree at thick slice (usual centroids), universal regime and thin slice while intermediate values were obtained by interpolation.}
\label{fig11}
\end{figure}

Interesting is the dependence of anisotropy level on the scale $R$. As shown in Table \ref{tab:absorpv}, in the presence of self-absorption structure function of centroids goes through different regimes as $R$ changes. Thus, the anisotropy level is expected to behave differently at different lags $R$. In particular, at small $R$ in thick-slice regime, the anisotropy level will be similar to the optically thin case of usual centroids, and this level will decrease with increase of $R$ as one passes through the universal regime towards thin-slice regime. This is demonstrated in Fig. \ref{fig11}.
Thus, the iso-correlation contours are more elongated at small $R$ and become more circular at large $R$.

Anisotropies in absorption line studies also show similar features. If absorption line is not saturated and the entire line is available for analysis, anisotropy level of centroid correlations will be similar to that of optically thin centroids. However, if absorption line is strong and only narrow portion of the wing is available for analysis, the anisotropy level will decrease reaching its minimum when the width of wing is essentially a delta function, and this minimum corresponds to the anisotropy level of thin-slice regime in VCA.

\section{Centroids and interferometric data}\label{sec:interfero}
Interferometric output can be used to directly measure spectra and anisotropy of the UVC statistics using the Fourier spatial components of the intensity distribution that are available with interferometers. With interferometer, one can obtain spectral intensity and multiply this with the line of sight velocity, and integrate this product over the entire frequency range to obtain centroid information. The study of centroids in Fourier domain is carried out in \citet{dutta2016visibility}, where potential of centroids to recover the turbulent velocity spectrum from radio interferometric observations is studied numerically. 

To study anisotropies using centroids, one can also use raw interferometric data. Anisotropic feature needs information of two dimensional maps, and therefore it is important to sample the fluctuations from different directions of the two dimensional $\bm{K}$ space. The turbulence anisotropy is manifested as the anisotropy of the power spectrum of interferometric data, and thus we write the power spectrum as
\begin{equation}\label{interferometerpower}
P(\bm{K})=\sum_{n=-\infty}^\infty P_n(K)\mathrm{e}^{-\mathrm{i}n\phi_K}~.
\end{equation}
Let us determine what information the measured $P_n$ coefficents can provide. Utilising Eqs. \eqref{centroidexp} and  \eqref{interferometerpower}, one can clearly write
\begin{equation}
P_n(K)=\int\mathop{\mathrm{d}^2\bm{R}} \mathcal{D}_n(R)\mathrm{e}^{-\mathrm{i}\bm{K}\cdot\bm{R}}~,
\end{equation}
which upon carrying out the angular integral yields
\begin{equation}
P_n(K)=\int\mathop{\mathrm{d}R}R\mathcal{D}_n(R)J_n(KR)~,
\end{equation}
where $J_n(KR)$ is the Bessel function of the first kind. In the case of $K\gg R^{-1}$, the multipole moment in Fourier space has an asymptotic form 
\begin{equation}\label{pninter}
P_n(K)=\mathcal{D}_n(\nu)\frac{2^{2+\nu}\Gamma\left(\frac{1}{2}(|n|-\nu+3)\right)}{\Gamma\left(\frac{1}{2}(|n|+\nu-1)\right)}(KS)^{3+\nu}~,
\end{equation}
where $\mathcal{D}_n(\nu)$ is the real space centroid structure moment after explicitly factoring out $R$ dependence, which for a constant density field is given by (c.f Eq. \eqref{centriodcoeff})
\begin{equation}
\mathcal{D}_n(\nu)=C_n(\nu)\sum_{p=-\infty}^\infty\hat{\mathcal{A}}_p\mathcal{W}_{n-p}~.
\end{equation}
It is clear from Eq. \eqref{pninter} that the centroid multipole moment in real space has one to one correspondence with that in the Fourier space, and the ratio of multipole moments in Fourier space is the same as that in real space. Therefore, by obtaining centroid anisotropy in Fourier space through interferometric data, one should be able to construct what that anisotropy corresponds to in real space. It should be stressed that this technique can also be used to obtain spectra of the velocity field. 

The study in \cite{esquivel2005velocity} suggests that MVCs might better reflect velocity statistics than UVCs. Thus, we believe that interferometric studies might also benefit by using MVCs. Since MVCs are defined only through the centroid structure function, MVCs in interferometric studies are equivalently defined through modified power spectrum by subtracting the product of velocity dispersion and power spectrum of column-density fluctuations. An elaborate technique equivalent to MVC would be to use fitting procedure with a supplied power spectrum of density fluctuations at sufficiently large $K$ to determine the velocity spectra. The advantage of the fitting procedure is one do not need to know velocity dispersion.

\section{Major findings of this paper}\label{sec:majorfinding}
The focus of this paper the description of unnormalised centroids for turbulence studies. The novel part of the paper is the description of the marked effects of absorption on the unnormalised centroids, and the description of centroid anisotropy based on the decomposition of MHD turbulence into different modes. As discussed in Sec. \ref{othick}, it is clear that one might not be able to  recover the usual centroids in the presence of absorption. As presented in Table \ref{tab:absorpv}, there is a range of scale where usual centroid asymptote can be obtained, and this range is set by the temperature of the turbulent cloud. Our results suggests that  centroids work better for warm clouds than for cold ones in the presence of absorption. This can also be understood from Eqs. \eqref{eq8} and \eqref{eq:rhos}. With increasing temperature (i.e. increasing $\beta_{\text{T}}$), the intensity $I_v$ of an optically thick medium matches closer to that of an optically thin medium. Therefore, the usual centroids can be recovered for larger $\beta_{\text{T}}$. Moreover, as shown in Fig. \ref{fig2}, there exists regimes of different scaling on $R$, and at some intermediate regime, the asymptote loses its dependence on velocity spectral index $\nu$, thus entering a  `universal' regime. Note that \cite{bertram2015centroid} carried out the numerical simulations to study gas dynamics in molecular clouds using velocity centroids. They observed a saturation of the spectral slope to value of $-3$, which corresponds to a scaling $\sim K^{-3}$ or equivalently $R$. Although their explanation was loosely based on VCA (\citetalias{lazarian2004velocity}), our results show that this saturation exists not only in VCA, but also in centroids. It is also important to note that in the presence of self-absorption, the column density gets affected the same way as velocity, and therefore, obtaining MVC is still possible. 

In Sec. \ref{abline}, we carried out the study on how one can apply centroids statistics to the absorption lines. It is clear from our analysis that study of turbulence with absorption lines is possible with centroids if one considers the centroids of logarithm of intensity instead of intensity. Our results clearly show that if one considers sufficiently small lag, then one can still obtain the usual centroids scaling $R^{1+\nu}$. The range of lag $R$ for which the usual centroid is obtained is set by width of the absorption window $\Delta$ and the thermal broadening $\beta_{\text{T}}$. For larger $\beta_{\text{T}}$, larger range of $R$ shows the usual centroids scaling. 

In terms of discussion on anisotropy, there are several important aspects that one can infer from our results. One of the important aspect is the issue of mode separation, i.e. identifying the composition of different MHD modes in a turbulent medium. With our result, it is clear that fast modes are easy to distinguish from Alfv\'en and slow due to two main reasons. First, fast modes show scaling in $R$ different from Alfv\'en and slow. Second, the quadrupole to monopole ratio of fast mode is positive while that of Alfv\'en and slow mode is negative.  This means that iso-correltion contours are elongated along sky-projected magnetic field direction for Alfv\'en and slow modes, and orthogonal to it for fast modes. Note that \citetalias{kandel2016extending} had similar findings on the distinctiveness of fast modes. With the present analysis, it is also clear that the separation of Alfv\'en and slow modes is challenging for two obvious reasons. First, the scaling on $R$ shown by the two modes is the same, and second both modes have the same anisotropic power spectrum, which implies similar level of anisotropy of centroid structure.

\section{Comparison between VCA, VCS and centroids}\label{vcacentroid}
The VCA introduced in \citetalias{lazarian2000velocity} provided a new foundation for studying velocity and density turbulence by studying the changes of spectral slope of intensity fluctuations within velocity slices of PPV cubes. It was shown in \citetalias{lazarian2000velocity} that by choosing a sufficiently thin slice, one might be able recover the velocity spectrum, whereas for sufficiently thick slice the velocity effects get washed away and only density spectrum can be recovered. The VCA was later extended in \citetalias{lazarian2004velocity} to account for the effects of self-absorption of emission lines. The results of \citetalias{lazarian2004velocity} suggests that in the presence of self-absorption, one might not be able to recover velocity spectrum, especially if the absorption is strong or the thermal broadening $\beta_{\text{T}}$ is larger than the dispersion of the velocity field. In fact, it was shown in \citetalias{lazarian2004velocity} that one might observe a universal spectrum $P(K)\sim K^{-3}$ at some intermediate lag $R$, which corresponds to the spectrum measured in a number of studies (see examples in \citealt{2009SSRv..143..357L}). The original formulation of VCA dealt only with power spectra, and this technique was further extended in \citetalias{kandel2016extending} to study the anisotropies induced by magnetic field in a plasma. This study showed how the anisotropy of an underlying turbulent field maps to anisotropy of intensity fluctuations, and showed how the level of anisotropy depends on the Alfv\'en Mach number as well as the angle between line of sight and mean magnetic field. 

Centroids are another powerful technique to study turbulence. We believe that our extension of centroids to study turbulence in a media with self-absorption, as well as for absorption-line significantly improves the value and power of this technique. In fact, we have also demonstrated how UVC can be used to study anisotropies of underlying turbulent field. 

We now compare the two techniques. The first important difference between the VCA and centroids is how thermal broadening $\beta_{\text{T}}$ affects them. While VCA cannot recover velocity spectrum at scales $R$ where $D_z(R)<\beta_{\text{T}}$, centroids still work in this regime. This means that VCA does not work well for subsonic turbulence unless we use emission lines from sub-dominant slow massive species. Centroids, on the other hand, are not reliable for supersonic turbulence (see \citealt{esquivel2007statistics}).  This is in contrast with VCA, which works well in this regime.

\begin{table*}
\centering
\caption{Comparison between centroids and VCA}
\begin{tabular}{l c c}  
\hline  
\emph{} & \emph{Velocity centroids} & \emph{VCA} \\
\hline
Uses & LOS velocity weighted intensity & Intensity\\
Best for & Subsonic turbulence & Supersonic turbulence\\
\hline
Scaling & $R^{1+\nu}$ & $R^{1-\nu/2}$ or $R^{1+\nu_\rho-\nu/2}$ (thin slice$^\dagger$)\\
 & & $R^{1+\nu_\rho}$ (thick slice$^{\dagger\dagger}$)\\
\hline
\end{tabular}
\begin{tabular}{l}
$^\dagger$For steep density $R^{1-\nu/2}$ dominates at small scales, while for shallow density $R^{1+\nu_\rho-\nu/2}$ dominates at small scales.\\
$\text{ }$ allows one to get information about velocity spectrum in thin slice limit.\\
$^{\dagger\dagger}$Velocity effects are washed away in the thick slice limit, and only density spectrum can be recovered.
\end{tabular}
\label{tab:comparison}
\end{table*}
In the presence of self-absorption, the VCA and the UVC share some similarities. As shown in Sec. \ref{othick}, in an optically thick medium one should expect a universal regime at some intermediate scales, where UVC lose their ability to trace the underlying velocity field, same as in the VCA. Similarly, for $\nu<2/3$ one should expect to see a thin slice asymptote $R^{1-\nu/2}$ at some larger scales even in centroids as in the VCA. All the expected scaling from velocity effects in the presence of self-absorption is presented in Table \ref{tab:absorpv} for both UVC and VCA. A brief comparison between VCA and UVC is also presented in Table \ref{tab:comparison}.

Although the spectral index of a velocity field is an important parameter, it does not provide a complete picture of an underlying turbulent field. In fact, for a magnetized plasma, anisotropies are important descriptors of the underlying field. We have shown in Sec. \ref{sec:mhdmodes} how the anisotropies of UVC structure function can be used to study the anisotropy of underlying turbulent field. This  provides a complimentary tool to study anisotropies. As shown in this paper and in \citetalias{kandel2016extending}, both VCA and UVC show similar traits in terms of anisotropy. First, the observed anisotropy in the thin slice regime in VCA and in UVC both show a clear dependence on both Alfv\'en Mach number and the angle between the magnetic field and the LOS. Second, both studies demonstrate that the anisotropy due to fast modes is opposite to that due to Alfv\'en and slow modes. However, to obtain complete understanding of anisotropies, one still needs to correctly obtain the angle between magnetic field and the LOS, the Alfv\'en Mach number and understand characteristic difference between Alfv\'en modes and slow modes. We believe that these understandings can be significantly enhanced by combining VCA and centroids and through model fittings.

Besides VCA and centroids, VCS is another powerful technique to study turbulence. Unlike VCA and centroids, VCS exclusively uses data along the velocity coordinate (in particular in the Fourier space), and one does not need to spatially resolve the scale of turbulence under study. A major advantage of VCS is that only few independent measurements are enough to obtain information about the underlying velocity field (Chepurnov \& Lazarian 2008). 

We stress the importance of using different techniques like VCA, VCS and centroids when studying  multi-phase ISM, e.g. H\,{\sevensize I} and H\,{\sevensize $\alpha$}. While VCA and VCS do not work well for gas components with large thermal broadening $\beta_{\text{T}}$ due to thermal dampening of the fluctuations\footnote{One strategy to use VCA and VCS in a hot medium is to use emission lines from heavy species like Fe.}, centroids work well for both hot and cold components at least as long as the turbulence is subsonic. Therefore, synergy of different techniques is advantageous, as these techniques have their own advantages. 

\section{Practical Aspects of Studies of Turbulence Using Centroids}\label{finiteres}
In this section, we discuss some of the practical aspects of studying turbulence with centroids. Firstly, when centroids are used to study turbulence is external galaxy, one must consider the contribution from galactic rotation to the LOS velocity, and thus to the centroids. Several methods are available to remove this contribution through the estimation of galactic rotation curve (see, for e.g, \citealt{miesch1994statistical}; \citealt{miesch1999velocity}; \citealt{dutta2016visibility}). In this paper, we have assumed that coherent motions such as rotations have been subtracted from the PPV data. 
Our approach has benefited from the formulation of centroids in PPV space which can
assume that such cleaning steps have been done before correlation studies.

Centroids are most suitable for sub-sonic turbulence. One of the candidates for objects of study for such turbulence is the clusters of galaxies. An example of a  cluster suitable for turbulence studies would be a nearby Coma cluster, where a potential cause of turbulence is an ongoing merger activity \citep{zuhone2016mapping}. The observational study of turbulence could be carried out with the data using new generation X-ray observatories like {\it Astro-H} \citep{takahashi2012astro}. Some of the practical aspects of studying turbulence has been   presented in \citet{zuhone2016mapping}, where they use normalised centroids to numerically test the ability of {\it Astro-H} to constrain various parameters (such as injection scale, Mach number). {\it Astro-H}, for instance, had capability to obtain line of sight velocities as well as velocity dispersion with a spectral resolution of order of km s$^{-1}$. Centroids are suitable when a good spatial coverage is available, otherwise a complimentary technique to study turbulence when such coverage is unavailable is VCS. VCS can be used with heavier species (like Fe) in clusters, for which the thermal broadening $\beta_\text{T}$ is small. The advantage of VCS is only few spectral measurements are enough to obtain turbulence spectra, and one does not even need a good spectral resolution to study turbulence.

X-ray observations are used to measure velocity and intensity profiles in such clusters, but X-ray observatories often have limited spectral resolution. Nonetheless, one should still be able to use centroids  as long as the dynamical range width $\Delta V_i$ of an instrument over which they are sensitive to signal is larger than the spread $\Delta V_0=\sqrt{D_z(S)+2\beta_{\text{T}}}$ of the signal. \citet[hereafter \citetalias{zhuravleva2012constraints}]{zhuravleva2012constraints} commented that with an energy resolution of $5$ eV and angular resolution of 1.7 arcmin, one should be able to measure the line profiles in Perseus cluster (where rms velocity of gas motions is $\sim 300$ km s$^{-1}$) with 90 per cent confidence level with the measurement time of $\sim 4\times 10^5$s.

Thermal broadening can put an important limitation on one's ability to obtain velocity dispersion, as the measured width of the line is affected by thermal motions of emitters. However, as noted in \citetalias{zhuravleva2012constraints}, thermal broadening is not important especially when one considers heavy species like Fe. It was discussed in \citetalias{zhuravleva2012constraints} that for instruments like {\it Astro-H} with energy resolution of 7 eV at 6.7 keV line width of Fe, and at gas motions of speed $\sim 400$ km s$^{-1}$, thermal broadening is unlikely to be significant.

An important limitation of an instrument is its finite spatial resolution. A finite diagram $\theta_0$ of an instrument introduces uncertainty $\delta R\sim \theta_0$ in the scales. Therefore, one should study centroids at scales $R>\theta_0$. Moreover, if the spatial resolution of the telescope is not good enough, then in the presence of self-absorption, one cannot see the turbulence induced statistics, but only the universal $\sim R$ regime.  To put this in perspective, we take a specific example of an instrument with 1.7$'$ angular resolution, which is used to study Perseus cluster, which is at a distance of 72 Mpc. With this resolution, only lags $R>20$ kpc can be studied to obtain turbulence spectra. 

The effect of instrumental noise has been discussed in detail in \citet[hereafter \citetalias{dickman1985largescale}]{dickman1985largescale} and \cite{miesch1994statistical}, where it was shown that actual correlations are underestimated in the presence of instrumental noise. \citetalias{dickman1985largescale} suggested to apply a constant multiplicative factor to account for noise contributions at non-zero lag. 

The studies of effects of noise in  \citetalias{dickman1985largescale} was carried out for normalised centroids. Here we apply their study for unnormalised centroids. In the presence of noise, the centroids structure function can be written as
\begin{equation}
\mathcal{D}_N(R)
=\mathcal{D}(R)+2\sigma_N^2~,
\end{equation}
where $\mathcal{D}(R)\equiv\left\langle\left[C(\bm{X}_1+\bm{R})-C(\bm{X}_1)\right]^2\right\rangle$, $\sigma_N^2$ is the variance due to noise. Since $\mathcal{D}(R)\propto R^n$, where $n>0$ is some index, it is clear that at extremely short scales, the centroid structure function can be corrupted by the noise. To estimate $\sigma_N$, we decompose the measured spectral intensity $I_{v,i}$ in a velocity channel $i$ into contributions from true emissions $I_{v,i}^s$ and contributions from noise $\delta I_{v,i}$, so that the centroids can be written as
\begin{equation}\label{noisecent}
C=\sum_{i=1}^NI_{v,i}^sv_i+\sum_{i=1}^N\delta I_{v,i}v_i~.
\end{equation}
The noise in two different spectroscopic channels are uncorrelated , and thus satisfies the constraints $\langle \delta I_{v,i}\delta I_{v,j}\rangle=\delta I^2$ for $i=j$ and 0 otherwise.  From Eq. \eqref{noisecent} it is clear that the error in centroids due to noise is $\delta C=\sum_{i=1}^N\delta I_{v,i}v_i$, and therefore the dispersion due to noise is given by
\begin{equation}\label{sigmanoise}
\sigma_N^2=\delta I^2\sum_{i=1}^Nv_i^2~.
\end{equation}
As shown in \citetalias{dickman1985largescale}, for large $N$, $\sum_{i=1}^Nv_i^2\approx \Delta^2N^3/12$, where $\Delta$ is the width of the spectroscopic channel. Using Eq. \eqref{sigmanoise}, one finally obtains
\begin{equation}
\sigma_N^2\approx\frac{\Delta^2N^3\delta I^2}{12}~.
\end{equation}
With this, the lags at which turbulence can be studied in the presence of noise is can be estimated by considering scales for which $\mathcal{D}(R)>2\sigma_N^2$. Restoring dimensional pre-factors in the expression for $\mathcal{D}(R)$, we obtain
\begin{equation}\label{rangenoise}
\frac{R}{S}>\left(\frac{\Delta^2N^3}{6D_z(S)}\frac{\delta I^2}{\bar{I}^2}\right)^{\frac{1}{1+\nu}}~,
\end{equation}
where $\bar{I}$ is the mean spectral intensity in a spectroscopic channel. In the case when background count per second in a spectral bin is given by $n_b$, the actual emission count per second in a spectral bin by $n_a$, one can write
\begin{equation}\label{sigmanoise1}
\frac{\delta I^2}{\bar{I}^2}=\frac{n_a+n_c}{n_a^2t}~,
\end{equation}
where $t$ is the measurement time. Eqs. \eqref{rangenoise} and \eqref{sigmanoise1} tell us that in the presence of noise, we can still study turbulence albeit not at extremely short scales. Moreover, the effect of noise can be reduced by increasing instrument time, as seen from Eq. \eqref{sigmanoise1}. It is important to note that to reduce the effects of noise, one should use a velocity window which is just wide enough to account for major signals.

\section{Discussion}\label{discuss}

\subsection{Foundations of the technique}

In this paper, we improve the understanding of centroids by studying unnormalised centroids in the presence of self-absorption, carrying out absorption line study and studying the effects of anisotropies in MHD turbulence. Unlike the past works, we explicitly use PPV space for absorption line study and study of self-absorption. This work shows the strength and usefulness of the PPV space formalism developed in \citetalias{lazarian2000velocity}.

In the view of modern understanding of MHD turbulence (see \citealt{beresnyak2015mhd} for a review), we study anisotropy of centroid correlation through explicit calculations in mode by mode basis. Theoretical and numerical research (\citetalias{goldreich1995toward},  \citealt{lithwick2001compressible}, \citealt{cho2002compressible, cho2003compressible}, \citealt{kowal2010velocity}) suggest that the MHD turbulence can be viewed as a superposition of the cascades of Alfven, slow and fast modes. The statistical properties of these cascades in the global frame of reference were obtained in \citetalias{lazarian2012statistical}  for the magnetic fields, while similar study was carried out in \citetalias{kandel2016extending} for the velocity field. In particular, \citetalias{kandel2016extending} studied turbulence anisotropies by making use of the VCA technique.  

\subsection{Areas of applicability of centroids}
One of the major advantage of centroids is their ability to study turbulence for subsonic regime using the main species, e.g. hydrogen. On the other hand, previous studies suggested that centroids are not reliable in the supersonic regime (\citealt{esquivel2005velocity}; \citealt{esquivel2007statistics}). This is distinct from VCA, which is applicable for supersonic turbulence. 

While thermal broadening does not affect the centroid statistics in the case of optically thin medium, as shown in Sec. \ref{othick}, thermal broadening might be an important parameter that determines whether or not we can recover the velocity statistics in the case of optically thick medium. 

Originally the centroids were developed for emission lines. However, our study in Sec. \ref{abline} extends UVC technique to study turbulence using absorption lines, which could be from the collection of point sources or from a spatially extended source. In fact, as we have shown in Sec. \ref{abline}, one may be able recover turbulence statistics at sufficiently small scales using absorption lines if one uses the centroids of logarithm of intensity. 

The prime objects of study using centroids are turbulence in diffuse ISM of the Milky Way and other galaxies and  in intergalactic gas in clusters of galaxies  using multi-wavelength single dish and interferometric measurements. The advantage of interferometric study is that one just needs a few measurements rather than restoring the entire PPV cubes to be able to perform the studies.

\subsection{Model assumptions}
In this paper, we adopted several model assumptions to make our analysis possible. One of the main assumptions is that the fluctuations are Gaussian. This assumption is satisfied by the velocity field to an appreciable degree (see \citealt{MoninYaglomLumley197504}). We do not make any assumption about Gaussianity of density field in the case when the turbulent medium is optically thin, as well as for absorption line studies, but we use Gaussian approximation for PPV space density to understand the main effects of self-absorption.  

While LE03 derived a general expression for centroid structure function keeping in mind that velocity and density might be correlated, we assumed that they are not. In fact, \cite{esquivel2007statistics} investigated the effects of density velocity correlation and showed that this correlation is not important if $\sigma_\rho/\rho_0\lesssim 1$. If this condition is not fulfilled, one should develop a model of density-velocity, perhaps basing on numerical simulation, correlation to retrieve full information from the centroids.  

Our analysis of centroid anisotropy was based on the decomposition of MHD turbulence into Alfv\'en, slow and fast modes. This decomposition is reasonable only when the coupling between the modes is marginal. \cite{cho2002compressible} showed that the degree of coupling between different modes to be moderate as long as the sonic Mach number is not very high. Since main regime where centroids are reliable is subsonic turbulence, this condition may not be restrictive for our purpose.

\subsection{New power of centroids}
This paper improves the usefulness of the UVC technique by providing an analytical description of the technique in the presence of self-absorption as well as for the absorption line study.  We believe that our study will be complimentary to the study in \citetalias{lazarian2004velocity}, where effects of self-absorption was studied in the context of VCA.  We also extend the ability of the centroids technique to study  magnetization of a media and direction of  the magnetic field, and explored the possibility of separating contributions of Alfv\'en, slow and fast modes. The separation of mode is important because different modes have different astrophysical impacts. As an example, Alfv\'en modes are essential for magnetic field reconnection (\citealt{lazarian1999reconnection},
see also \citealt{lazarian2015turbulent} and ref. therein), superdiffusion of cosmic rays (\citealt{lazarian2014superdiffusion}), etc. On the other hand, fast modes are important for resonance scattering of cosmic rays (\citealt{yan2002scattering}). The possible ability of centroids to obtain the relative contribution of these different modes complements this ability for the technique introduced in \citetalias{lazarian2012statistical}  and \citetalias{0004-637X-818-2-178} for synchrotron data and in \citetalias{kandel2016extending} for spectroscopic data. 

\subsection{Centroids and other techniques}
In this paper, we studied centroids by extensively making use of PPV space formalism. We have also discussed and compared centroids with the VCA and VCS, the techniques that were developed also by using PPV space formalism.  Centroids have been used to analytically study anisotropies in this paper, while anisotropies were studied using VCA in \citetalias{kandel2016extending}. Another technique called principal component analysis (PCA, see \citealt{brunt2002interstellar}) can also be used to study turbulence anisotropies. However, unlike the centroids and VCA, it is not easy to quantify PCA using PPV data. Nevertheless, recent studies have shown the sensitivity of PCA to the phase information (\citealt{0004-637X-818-2-118}), although the trend is not yet clear.

Another important technique to study turbulence using velocity slice of PPV space is the spectral correlation function (SCF, see \citealt{rosolowsky1999spectral}). The SCF is very similar to VCA if one removes the adjustable parameters from SCF.  In fact, both SCF and VCA measure correlations of intensity in velocity slices of PPV, but the SCF treats outcomes empirically. There also exist numerous techniques identifying and analyzing clumps and shells in PPV (see \citealt{houlahan1992recognition}; \citealt{williams1994determining}; \citealt{stutzki1990high}; \citealt{pineda2006complete}; \citealt{ikeda2007survey}). 

Besides the VCA and centroids, there are also some other techniques to study sonic and Alfv\'en Mach numbers. Some of these techniques  include so called Tsallis statistics (see \citealt{esquivel2010tsallis}; \citealt{tofflemire2011interstellar}), bi-spectrum (see \citealt{burkhart2009density}), genus analysis (see \citealt{chepurnov2008topology}), etc. Using different available techniques allows one to obtain a comprehensive picture of MHD turbulence.
 
\section{Summary}\label{summ}
On the basis of the analytical theory of the intensity fluctuations in the PPV space developed in the earlier works (\citetalias{lazarian2000velocity}; \citetalias{lazarian2004velocity};  \citetalias{lazarian2008studying}; \citetalias{kandel2016extending}) we provided the analytical description of the statistics of fluctuations measured by velocity centroids. Our definition of centroids follows that in \citetalias{lazarian2003statistics} and differs from the traditionally used by the absence of the normalization by intensity. While the normalization makes analytical studies really prohibitive, it was shown in numerical studies in \cite{esquivel2005velocity} to be of no significance for restoring underlying velocity statistics. Therefore, we use unnormalized velocity centroids \citep{esquivel2005velocity} (see Table \ref{tab:centroidstype}). Our results can be briefly summarized as follows

\begin{itemize}

\item We proved the complementary nature of turbulence studies with velocity centroids and the Velocity Channel Analysis  (VCA). Both techniques can measure turbulence spectra and anisotropy. With VCA one can explore supersonic turbulence where numerical studies show that centroid technique is unreliable. At the same time centroids deliver turbulence statistics for subsonic turbulence. The VCA can study subsonic turbulence only using heavier species, e.g. metals in hydrogen gas. The anisotropies revealed by centroids are of higher amplitude. Nevertheless, a more accurate separation of density and different MHD mode contribution is possible with the VCA through varying the width of the velocity channel.

\item Analytical expressions for UVC structure function are obtained in the presence of emission lines in the presence of self-absorption as well as for the absorption lines. In the presence of self- absorption new scalings of fluctuations measured by UVC at different scales are reported.  Similar to the VCA, for a range of scales, the UVCs correlations are shown to exhibit universal scaling, thus losing the information of the velocity spectra.

\item For absorption line, we suggest to construct UVC as LOS velocity weighted logarithm of intensity, and focus on the wings where absorption lines is not saturated. We termed thus defined centroid as restricted velocity centroid (RVC) (see Eq. \eqref{sharpdef}) and show that at sufficiently small lags it exhibits the usual centroids scaling. RVCs open a new way of probing astrophysical turbulence. Both turbulence statistics and turbulence anisotropies can be studied with RVCs.

\item Analytical expression of UVC structure function when turbulence is sub-Alfv\'enic and thus anisotropic is derived. This expression is used to study the anisotropy arising from three MHD modes: Alfv\'en, fast and slow. It is shown that the quadrupole to monopole ratio of fast mode is positive, while it is negative for Alfv\'en and slow mode. In other words, iso-correlation contours are elongated along sky-projected magnetic field direction for Alfv\'en and slow modes, and orthogonal to it for fast modes, which is same as what VCA predicts (see \citetalias{kandel2016extending}).

\item Self-absorption of the radiated emission does not preclude anisotropy studies with UVCs, MVCs and RVCs. Our study suggests that at sufficiently short lag $R$, one will observe the same level of anisotropy as that of optically thin centroids, while at larger scales the anisotropy level decreases to the level of thin-slice regime of VCA.

\item Interferometer output can be used to directly measure spectra and anisotropy of the UVC statistics using the Fourier spatial components of the intensity distribution that are available with interferometers. This considerably simplifies such measurements, as there is no necessity to restore the image of the turbulent volume. Instead, having a few of these Fourier components is sufficient to get both spectra and anisotropy of turbulence.

\end{itemize}

\section{Acknowledgements}
D~.K. and D.~P. thank the Institut Lagrange de Paris, a LABEX funded by the ANR (under reference ANR-10-LABX-63) within the Investissements d'Avenir programme under reference ANR-11-IDEX-0004-02. A.~L. acknowledges the NSF grant AST 1212096 and Center for Magnetic Self Organization (CMSO). He also acknowledges Institut Astrophysique du Paris and Institut Lagrange de Paris for hospitality during his visit. 


\footnotesize{
\nocite{*}
\bibliographystyle{mnras}
\bibliography{Centroids}
}
\bsp
\label{lastpage}
\end{document}